\documentclass[aps,prd,superscriptaddress,10pt,showpacs,notitlepage,  
nofootinbib
]{revtex4-1}
\usepackage{bm}
\usepackage{amsmath,amssymb,amsthm}
\usepackage{latexsym,graphicx,color,subfigure}
\usepackage{enumerate}
\usepackage{amssymb}
\usepackage{hyperref}
\usepackage{mathtools}

\usepackage{graphicx}
\graphicspath{ {images/} }

\usepackage{color}

\newcommand{\be}{\begin{equation}}
\newcommand{\ee}{\end{equation}} 
\newcommand{\beq}{\begin{eqnarray}}
\newcommand{\eeq}{\end{eqnarray}}

\newcommand{\bea}{\begin{eqnarray}}
\newcommand{\eea}{\end{eqnarray}}

\def\tc{\textcolor{red}}
\def\tc2{\textcolor{blue}}
 
\newcommand{\drm}{\mathrm{d}}
\renewcommand{\vec}[1]{\boldsymbol{#1}}

\allowdisplaybreaks[3]
\def\simge{\mathrel{
   \rlap{\raise 0.511ex \hbox{$>$}}{\lower 0.511ex \hbox{$\sim$}}}}
\def\simle{\mathrel{
   \rlap{\raise 0.511ex \hbox{$<$}}{\lower 0.511ex \hbox{$\sim$}}}}
\def\bigs{\mathrel{
   \rlap{\raise 0.531ex \hbox{$>$}}{\lower 0.531ex \hbox{$<$}}}}

\usepackage[normalem]{ulem}

\renewcommand\sout{\bgroup \color{blue} \ULdepth=-.5ex \ULset}
\def\tc{\color{red}}

\begin{document}
\title{Reexamining Ginzburg-Landau theory for neutron $^{3}P_{2}$ superfluidity in neutron stars}
\author{Shigehiro Yasui}
\email{yasuis@keio.jp}
\affiliation{Department of Physics $\&$ Research and Education Center for Natural Sciences,\\ Keio University,Hiyoshi 4-1-1, Yokohama, Kanagawa 223-8521, Japan}
\author{Chandrasekhar Chatterjee}
\email{chandra@phys-h.keio.ac.jp}
\affiliation{Department of Physics $\&$ Research and Education Center for Natural Sciences,\\ Keio University,Hiyoshi 4-1-1, Yokohama, Kanagawa 223-8521, Japan}
\author{Michikazu Kobayashi}
\email{michikaz@scphys.kyoto-u.ac.jp}
\affiliation{Department of Physics, Kyoto University, Oiwake-cho, Kitashirakawa, Sakyo-ku, Kyoto 606-8502, Japan}
\author{Muneto Nitta}
\email{nitta(at)phys-h.keio.ac.jp}
\affiliation{Department of Physics $\&$ Research and Education Center for Natural Sciences,\\ Keio University,Hiyoshi 4-1-1, Yokohama, Kanagawa 223-8521, Japan}
\date{\today}
\begin{abstract}
The Ginzburg-Landau (GL) effective theory is a useful tool to 
study a superconductivity or superfluidity near the critical temperature, 
and usually the expansion up to 
the 4th order in terms of order parameters is sufficient for the description of the second-order phase transition.
In this paper, we discuss the GL equation for the neutron $^{3}P_{2}$ superfluidity relevant for interior of neutron stars.
We derive the GL expansion up to the 8th order 
in the condensates
and find that this order is necessary for the system to have 
the unique ground state, 
unlike the ordinary cases. 
Starting from the $LS$ potential, which provides the dominant attraction between two neutrons at the high density,
we derive the GL equation in the path-integral formalism, where the auxiliary field method and the Nambu-Gor'kov representation are used.
We present the detailed description for the trace calculation necessary in the derivation of the GL equation.
As numerical results, we show the phase diagram of the neutron $^{3}P_{2}$ superfluidity on the plane spanned by the temperature and  magnetic field, and find that the 8th order terms lead to a first-order phase transition, whose existence 
was predicted in the Bogoliubov-de Gennes equation 
but has not been found thus far 
within the framework of the GL expansion up to the 6th order.
The first-order phase transition will affect the interior structures inside the neutron stars.
\end{abstract}
\maketitle


\section{Introduction}

Neutron stars are one of the important astrophysical objects providing us laboratories of nuclear physics (see Refs.~\cite{Graber:2016imq,Baym:2017whm} for recent reviews). 
The extreme environments in the neutron stars, such as high density state, rapid rotation, strong magnetic field, strong gravitational field, and so on, lead to interesting questions about the unconventional states of nuclear systems.
For example, it is considered that there can exist a various kind of matter phases inside the neutron stars:
neutron rich gas and crusts at the surface, and 
neutron superfluidity, hyperon matter, $\pi$, $K$ condensates, quark matter, etc., in the inside.
Those states can be accessible from observations of the mass-radius relation, time-evolution of the surface temperature, neutrino emissions, and so on.
The research of the equation of state is still an open question, and many efforts have been devoted to understand the massive neutron stars whose masses are almost twice as large as the solar mass~\cite{Demorest:2010bx,Antoniadis1233232}.
One of the most recent observational developments were provided by the gravitational waves from 
the binary neutron star merger~\cite{TheLIGOScientific:2017qsa}. 
This finding opens a new era of researches of neutron stars in terms of the multimessenger astronomy including the microscopic physics, such as nuclear and elementary particle physics.

Among the various phases in the neutron stars,
the research of
 the neutron superfluidity has bridged the connection between the theoretical knowledge and the astrophysical observations (see Refs.~\cite{Chamel2017,Haskell:2017lkl,Sedrakian:2018ydt} for recent reviews).
It was discussed that the neutron superfluidity affects relaxation time after pulsar glitches (sudden speed-up events of neutron-star rotation)~\cite{Reichely1969}.
It was also discussed that the neutron superfluidity is related to rapid cooling by neutrino emissions, which is known as the modified Urca process~\cite{Yakovlev:1999sk}.
It was proposed that pulsar glitch phenomena can be explained by pinning of large amount of superfluid vortices 
\cite{Anderson:1975zze}. 
From a more microscopic point of view, 
it is worthwhile to remind us that nuclear forces can supply the attractions in different channels for the baryon number density from low density to high density~(see Ref.~\cite{Dean:2002zx} for a recent review). 
Migdal considered the $^{1}S_{0}$ channel as the most attractive interaction at low density~\cite{Migdal:1960}.
However, the $^{1}S_{0}$ channel becomes repulsive at higher densities due to the strong repulsion at short ranges of the nuclear force~\cite{1966ApJ...145..834W}.
At such high density, 
 the attractive channel is given by the $^{3}P_{2}$ channel from the $LS$ potential.\footnote{We notice that the $^{3}P_{0}$ and $^{3}P_{1}$ channels are repulsive.
}
This attraction
 induces the neutron $^{3}P_{2}$ superfluidity as it was obtained by solving the gap equation~\cite{Tabakin:1968zz,Hoffberg:1970vqj,Tamagaki1970,Takatsuka1971,Takatsuka1972,Amundsen:1984qc,Takatsuka:1992ga,Baldo:1992kzz,Elgaroy:1996hp,Khodel:1998hn,Baldo:1998ca,Khodel:2000qw,Zverev:2003ak,Maurizio:2014qsa,Bogner:2009bt,Srinivas:2016kir}.
Thus,
neutron $^{3}P_{2}$ superfluidity has been studied as one of the interesting states of the nuclear matter inside the neutron stars.
Notice that, precisely to say, the neutron $^{3}P_{2}$ superfluidity exists with a small fraction of superconducting protons and normal electrons.
As an astrophysical application, the rapid cooling time in neutron stars in Cassiopeia A was studied in terms of the the neutron $^{3}P_{2}$ superfluidity~\cite{Heinke2010,Shternin2011,Page:2010aw}, although the existence of the $^{3}P_{2}$ superfluidity is still elusive. 
It is also important to mention that, while the $^{1}S_{0}$ pairing is fragile in 
the presence of strong magnetic fields, the $^{3}P_{2}$ pairing is not.
Therefore, 
one of possibilities to observe $^{3}P_{2}$ superfluidity is 
in magnetars, accompanied by large amount of magnetic fields.
The magnetic field of magnetars reaches about $10^{15}$ G at the surface and may reach even $10^{18}$ G in the inside, 
and they are interesting astrophysical objects to study the neutron $^{3}P_{2}$ superfluidity under the superstrong magnetic field.\footnote{The origin of the strong magnetic fields was studied in terms of several mechanisms such as spin-dependent interactions between neutrons~\cite{Brownell1969,RICE1969637,Silverstein:1969zz,Haensel:1996ss},
the pion domain wall~\cite{Eto:2012qd,Hashimoto:2014sha}, the spin polarization in quark-matter core~\cite{Tatsumi:1999ab,Nakano:2003rd,Ohnishi:2006hs}, and so on. However, 
in recent years, the many-body calculation leads to a negative result for the realization of strong magnetic fields~\cite{Bordbar:2008zz}.}

The pairing in the neutron $^{3}P_{2}$ superfluidity is provided by the strong $LS$ potential between two neutrons as mentioned above.
One of the particular features of the $^{3}P_{2}$ superfluidity is the richness of the symmetries of the condensate, 
which stems from the variety of the pairing patterns due to the various combinations between the relative angular momentum and the total spin.
The ground state has the symmetry breaking from the $\mathrm{U}(1)_{B} \times \mathrm{SO}(3)_{S} \times \mathrm{SO}(3)_{L} \times T \times P$ symmetry ($B$ for baryon number, $S$ for spin rotation, $L$ for spatial rotation, $T$ for time-reversal symmetry, and $P$ for parity symmetry)~\cite{Fujita1972,Richardson:1972xn,Sauls:1978lna,Muzikar:1980as,Sauls:1982ie,Vulovic:1984kc,Masuda:2015jka,Masuda:2016vak}.
It was discussed that, as a consequence of the symmetry breaking, low-energy excitations affect the cooling process by neutrino emission~\cite{Bedaque:2003wj,Bedaque:2012bs,Bedaque:2013fja,Bedaque:2014zta,Leinson:2009nu,Leinson:2010yf,Leinson:2010pk,Leinson:2010ru,Leinson:2011jr,Leinson:2012pn,Leinson:2013si,Leinson:2014cja}.
We comment that the cooling process is also related to quantum vortices~\cite{Shahabasyan:2011zz}.
Recently the studies of the topological properties of the neutron $^{3}P_{2}$ superfluidity has been conducted:
topological superfluidity and 
gapless Majorana fermions on the boundary of $^{3}P_{2}$ superfluids~\cite{Mizushima:2016fbn}, 
a quantized vortex~\cite{Masuda:2015jka}, 
a soliton on it~\cite{Chatterjee:2016gpm} 
and a half-quantized non-Abelian vortex~\cite{Masuda:2016vak}.
Those studies were stimulated from the analogous states in condensed matter systems, 
such as $D$-wave superconductivity~\cite{Mermin:1974zz}, 
$P$-wave superfluidity in $^{3}$He liquid~\cite{vollhardt2013superfluid,Volovik:2003fe}, 
chiral $P$-wave superconductivity, e.g., in Sr$_2$RuO$_4$~\cite{RevModPhys.75.657}, 
spin-2 Bose-Einstein condensates~\cite{2010arXiv1001.2072K}, and so on.
Thus, the research of the neutron $^{3}P_{2}$ superfluidity shares common interests in various condensed matter systems 
as an interdisciplinary subject.

The basic equation for the neutron $^{3}P_{2}$ superfluidity is the Bogoliubov--de-Gennes (BdG) equation~
\cite{Tabakin:1968zz,Hoffberg:1970vqj,Tamagaki1970,Takatsuka1971,Takatsuka1972,Amundsen:1984qc,Takatsuka:1992ga,Baldo:1992kzz,Elgaroy:1996hp,Khodel:1998hn,Baldo:1998ca,Khodel:2000qw,Zverev:2003ak,Maurizio:2014qsa,Bogner:2009bt,Srinivas:2016kir}.
Recently it was applied
 to investigate phase structures and topological properties in neutron $^{3}P_{2}$ superfluidity~\cite{Mizushima:2016fbn}.
Around the critical temperature, the BdG equation can be reduced to the Ginzburg-Landau (GL) equation as the low-energy effective theory~\cite{Fujita1972,Richardson:1972xn,Sauls:1978lna,Muzikar:1980as,Sauls:1982ie,Vulovic:1984kc,Masuda:2015jka,Masuda:2016vak}.
The GL equation is obtained by integrating-out the fermionic degrees of freedom, and it is expressed in terms of the pairing gap behaving as a bosonic degree of freedom.
It is known that, at the weak-coupling limit, the ground state of the neutron $^{3}P_{2}$ superfluidity 
is in the nematic phase, {\it i.e.}, uniaxial nematic (UN) or biaxial nematic (BN) phase depending on the temperature and magnitude of the magnetic field~\cite{Sauls:1978lna}.
The UN phase has an unbroken U(1) symmetry, while
the BN phase is furthermore classified into two types according to symmetries, 
{\it i.e.}, D$_{2}$-BN and D$_{4}$-BN phases,
having an unbroken dihedral symmetries,
D$_{2}$ and D$_{4}$, respectively. 
The UN phase is favored at zero magnetic field, 
while the BN phases are favored in the presence of finite magnetic fields~\cite{Masuda:2015jka,Mizushima:2016fbn},
among which 
the D$_{4}$-BN phase is favored 
above the critical magnetic field relevant for magnetars.
Since the GL equation is much easier to deal with than the BdG equation,
it has been applied to various problems although its use is limited to the vicinity of the critical temperature.
For instance, it has been applied to study 
vortex structures of neutron $^{3}P_{2}$ superfluidity~\cite{Richardson:1972xn,Muzikar:1980as,Sauls:1982ie}, 
inside whose core spontaneous magnetization was found~\cite{Sauls:1982ie,Masuda:2015jka,Masuda:2016vak,Chatterjee:2016gpm}. 
The GL equation is given in terms of the expansion series of the order parameter of the neutron $^{3}P_{2}$ superfluidity.
Thus, it is an important question what order in the GL expansion 
is enough to capture physics of the neutron $^{3}P_{2}$ superfluidity.
The 4th order terms are enough in usual cases but it is not so for the case of the neutron $^{3}P_{2}$ superfluidity 
as explained below.
There is a continuous degeneracy in the ground states in the absence of the magnetic field 
as long as we consider the terms up to the 4th order for the order parameter.
In other words, there is no energy difference among the UN, D$_{2}$-BN, and D$_{4}$-BN phases,
and therefore we cannot determine the unique ground state.\footnote{At this order, there happens to exist an $\mathrm{SO}(5)$ symmetry as an extended symmetry of the potential term, which is absent in the Hamiltonian, and spontaneous breaking of such the symmetry leads a quasi-Nambu-Goldstone mode~\cite{Uchino:2010pf}.
}
When we include the 6th order terms, only one phase is realized 
as the most stable state among them, 
{\it i.e.}, the UN phase for zero magnetic field and the BN phases for finite magnetic field as mentioned above.
The effect of the strong magnetic field beyond the leading-order term was analyzed in the phase diagram and the thermodynamic properties~\cite{Yasui:2018tcr,Yasui:2019tgc}.
Such information will be useful to study the neutron $^{3}P_{2}$ superfluidity in magnetars.
However, there is a serious drawback of the expansion up to the 6th order. 
The 6th order terms make the system unstable; 
There is the instability for large values of the order parameter 
and therefore the energy is not bounded from below.

In the present paper, 
to resolve this problem,
we study the 8th order term of the order parameter 
which has not been studied so far.
We find that the GL expansion up to the 8th order is the minimum order to uniquely determine the stable ground states 
(including the case of zero magnetic fields).
We then investigate the phase diagram of the neutron $^{3}P_{2}$ superfluidity at this order,
and find the existences of a 1st order phase transition line between the D$_2$-BN and D$_4$-BN phases 
and a tricritical point separating the 1st and 2nd order phase transition lines.
Although a first-order line was previously predicted in the BdG equation under the quasiclassical approximation 
\cite{Mizushima:2016fbn}, it was not found in the GL expansion up to the 6th order.
Therefore, the GL expansion up to the 8th order is the minimum order to capture the essential physics of 
phase transitions in the $^{3}P_{2}$ superfluidity.
As such, the first-order phase transition may affect the interior structures inside the neutron stars.
We also present the detailed description about the derivation of the GL equation in the path-integral formalism, in which the auxiliary field method and the Nambu-Gor'kov representation are introduced.

This paper is organized as follows.
In Sec.~\ref{sec:Lagrangian}, we introduce the $LS$ interaction between two neutrons, and give the tensor-type interaction relevant to the neutron $^{3}P_{2}$ superfluidity.
In Sec.~\ref{sec:GL_eq}, we give the detailed description for deriving the GL equation with the 8th order term by adopting the path-integral formalism.
In Sec.~\ref{sec:phase_diagram}, we show the numerical results for the phase diagram of the neutron $^{3}P_{2}$ superfluidity, and discuss the effect of 
the 8th order term on the change of the phase diagram.
The final section is devoted to our conclusion.
In Appendix~\ref{sec:trace_calculation},
we present some details of the trace calculations used in the derivation of the GL equation.
In Appendix~\ref{sec:symmetries}, we summarize symmetries of the $^3$P$_2$ order parameters.

\section{Interaction Lagrangian}
\label{sec:Lagrangian}

We introduce a nonrelativistic two-spinor field $\varphi(t,\vec{x})=(\varphi_{\uparrow}(t,\vec{x}),\varphi_{\downarrow}(t,\vec{x}))^{t}$ for a neutron.
We consider the Lagrangian as a sum of the kinetic term and the the term for the interaction with the magnetic field,
\begin{eqnarray}
 {\cal L}_{0}
=
 \varphi^{\dag}(t,\vec{x}) \biggl( i\partial_{t} + \frac{\vec{\nabla}^{2}}{2m} + \mu + \vec{\mu}_{n} \!\cdot\! \vec{B} \biggr) \varphi(t,\vec{x}),
\end{eqnarray}
with the neutron mass $m$ and the chemical potential $\mu$.
Notice that the interaction potential between the spin and the magnetic field is given by $\displaystyle V_{m} = -\varphi^{\dag} \vec{\mu}_{n} \!\cdot\! \vec{B} \varphi$.
Here
$\displaystyle \vec{\mu}_{n} =  - \gamma_{n}\hbar \vec{s}$ with the spin operator $\vec{s}=\vec{\sigma}/2$ with the Pauli matrices $\vec{\sigma}=(\sigma^{1},\sigma^{2},\sigma^{3})$
is the magnetic moment and $\gamma_{n}=1.2 \times 10^{-13}$ MeV/T ($1 \, \mathrm{T}=10^{4} \, \mathrm{G}$).

We consider the $LS$ interaction between two neutrons as the dominant force ~\cite{Tabakin:1968zz,Hoffberg:1970vqj,Tamagaki1970,Takatsuka1971,Takatsuka1972,Amundsen:1984qc,Takatsuka:1992ga,Baldo:1992kzz,Elgaroy:1996hp,Khodel:1998hn,Baldo:1998ca,Khodel:2000qw,Zverev:2003ak,Maurizio:2014qsa,Bogner:2009bt,Srinivas:2016kir}.
In the center-of-mass frame, we consider the scattering process $(\vec{p},j)+(-\vec{p},l) \rightarrow (\vec{q},i)+(-\vec{q},k)$ for the initial momentum $\vec{p},-\vec{p}$ and spin $j,l=\uparrow,\downarrow$ and the final momentum $\vec{q},-\vec{q}$ and spin $i,k=\uparrow,\downarrow$.
Then, the interaction Lagrangian of the $LS$ interaction is provided by
\begin{eqnarray}
 {\cal L}_{\mathrm{int}}
= 
   \sum_{i,j,k,l}
   \varphi_{\vec{q}i}^{\ast} \varphi_{\vec{p}j} \bigl(V_{LS}\bigr)_{ijkl} \varphi_{-\vec{q}k}^{\ast} \varphi_{-\vec{p}l},
\label{eq:LS_potential_momentum}
\end{eqnarray}
where we define the $LS$ operator defined by
\begin{eqnarray}
  \bigl(V_{LS}\bigr)_{ijkl} = i 2G'  
   (\vec{S})_{ijkl} \!\cdot\! \bigl( \vec{p} \times \vec{q} \bigr),
\end{eqnarray}
with the coupling constant $G'>0$ and the spin operator $(\vec{S})_{ijkl}=(\vec{s})_{ij}\delta_{kl}+\delta_{ij}(\vec{s})_{kl}$ with $i,j,k,l=\uparrow,\downarrow$.
For the later calculation, we transform the $LS$ interaction \eqref{eq:LS_potential_momentum} in the momentum space to another form in the real space.
For this purpose, adopting the Fourier transformation
\begin{eqnarray}
  \varphi_{\vec{p}} = \int \drm^{3}\vec{x} \, e^{-i\vec{p}\cdot\vec{x}} \varphi(t,\vec{x}),
\label{eq:nucleon_field_FT_2}
\end{eqnarray}
then we obtain
\begin{eqnarray}
  \int \frac{\drm^{3}\vec{p}}{(2\pi)^{3}} \frac{\drm^{3}\vec{q}}{(2\pi)^{3}} {\cal L}_{\mathrm{int}}
&=&
  -G' 
  \int \drm^{3}\vec{x} \, \drm^{3}\vec{y} \,
  \biggl(
         \frac{2}{3} S^{\dag}(t,\vec{x}) S(t,\vec{y})
     + \frac{1}{2} \sum_{a} V^{a\dag}(t,\vec{x}) V^{a}(t,\vec{y})
     - \sum_{a,b} T^{ab\dag}(t,\vec{x})T^{ab}(t,\vec{y})
  \biggr),
\label{eq:int_relation1a}
\end{eqnarray}
where we have introduced scalar, vector and tensor fields by
\begin{eqnarray}
  S(t,\vec{x}) &\equiv& \sum_{a} \phi^{aa}(t,\vec{x}), \label{eq:scalar_def} \\
  V^{a}(t,\vec{x}) &\equiv& \sum_{b,c} \epsilon^{abc} \phi^{bc}(t,\vec{x}), \label{eq:vector_def} \\
  T^{ab}(t,\vec{x}) &\equiv& \frac{1}{2} \Bigl( \phi^{ab}(t,\vec{x}) + \phi^{ba}(t,\vec{x}) \Bigr) - \frac{1}{3} \delta^{ab} \sum_{c} \phi^{cc}(t,\vec{x}),
\label{eq:tensor_def}
\end{eqnarray}
with $a,b,c=1,2,3$. 
Notice that the tensor field is symmetric and traceless.
For the short notation, we define for the bilinear field for two neutrons
\begin{eqnarray}
    \phi^{ab}(t,\vec{x})
\equiv
 - \varphi(t,\vec{x})^{t} \Sigma^{a\dag} \bigl( \nabla^{b}_{x} \varphi(\tau,\vec{x}) \bigr),
\label{eq:phi_def}
\end{eqnarray}
with the spin operators defined by
\begin{eqnarray}
  \Sigma^{0} \equiv i\sigma^{2}, \hspace{1em}
  \Sigma^{1} \equiv i\sigma^{1}\sigma^{2} = -\sigma^{3}, \hspace{1em}
  \Sigma^{2} \equiv i\sigma^{2}\sigma^{2} = i \vec{1}, \hspace{1em}
  \Sigma^{3} \equiv i\sigma^{3}\sigma^{2} = \sigma^{1},
\label{eq:Sigma_def}
\end{eqnarray}
and the spatial derivative $\nabla^{b}_{x}=\partial/\partial x_{b}$.
Here $\varphi^{t}\Sigma^{0}\varphi$ is invariant as the scalar and $\varphi^{t}\vec{\Sigma}\varphi$ with $\vec{\Sigma}=(\Sigma^{1},\Sigma^{2},\Sigma^{3})$ transforms as a vector.
This is checked by the spin rotation of $\varphi$.
We notice that Eq.~\eqref{eq:int_relation1a} is the nonlocal interaction.
For simplicity, we regard this interaction as the local one by replacing the nonlocal operator $X(t,\vec{x})=S(t,\vec{x})$, $V^{a}(t,\vec{x})$, and $T^{ab}(t,\vec{x})$ in the following way:
\begin{eqnarray}
 G' X^{\dag}(t,\vec{x})X(t,\vec{y}) \rightarrow G \, X^{\dag}(t,\vec{x})X(t,\vec{x}) \delta^{(3)}(\vec{x}-\vec{y}),
\end{eqnarray}
where we have introduced a new coupling constant $G(>0)$ instead of the original one $G'$.
The mass dimension of $G$ is different from that of $G'$.
Then, Eq.~\eqref{eq:int_relation1a} is reduced to
\begin{eqnarray}
&&
  -G' 
  \int \drm^{3}\vec{x} \, \drm^{3}\vec{y} \,
  \biggl(
         \frac{2}{3} S^{\dag}(t,\vec{x}) S(t,\vec{y})
     + \frac{1}{2} \sum_{a} V^{a\dag}(t,\vec{x}) V^{a}(t,\vec{y})
     - \sum_{a,b} T^{ab\dag}(t,\vec{x})T^{ab}(t,\vec{y})
  \biggr)
\nonumber \\
&\rightarrow&
  -G
  \int \drm^{3}\vec{x} \,
  \biggl(
         \frac{2}{3} S^{\dag}(t,\vec{x}) S(t,\vec{x})
     + \frac{1}{2} \sum_{a} V^{a\dag}(t,\vec{x}) V^{a}(t,\vec{x})
     - \sum_{a,b} T^{ab\dag}(t,\vec{x})T^{ab}(t,\vec{x})
  \biggr).
\label{eq:int_relation1}
\end{eqnarray}
Because the scalar and vector channels are repulsive for $G>0$ and these two channels will be irrelevant to the neutron superfluidity in a direct manner,
we will neglect those two terms in the following discussions.\footnote{We keep it in mind that the magnetic field may cause the mixing between the scalar, vector, and tensor channels, when the strength of the magnetic field is strong enough.
If such mixing occurs, then the tensor interaction may be more enhanced through the virtual processes with the scalar and vector channels.}
From the above consideration, we obtain the full Lagrangian relevant to the tensor interaction
\begin{eqnarray}
 {\cal L}
=
 \varphi^{\dag}(t,\vec{x})\biggl( i\partial_{t} + \frac{\vec{\nabla}^{2}}{2m} + \mu + \vec{\mu}_{n} \!\cdot\! \vec{B} \biggr) \varphi(t,\vec{x})
+ G \sum_{a,b} T^{ab\dag}(t,\vec{x})T^{ab}(t,\vec{x}),
\label{eq:Lagrangian_m}
\end{eqnarray}
as the basic Lagrangian for the neutron $^{3}P_{2}$ superfluidity.
In the next section, we will derive the GL free energy density from this Lagrangian.

\section{Derivation of the Ginzburg-Landau equation}
\label{sec:GL_eq}

\subsection{Auxiliary field method}

Based on the Lagrangian \eqref{eq:Lagrangian_m}, we consider the generating functional.
For the convenience of the calculation, we introduce the Wick rotation for the real time $t$ on the complex plane, and consider the imaginary time $\tau$ by replacing $it$ by $\tau$ in the Euclidean space and treat $\tau$ as a real parameter.
Because we are interested in the finite temperature, we restrict the range of $\tau$ to the periodic range, $\tau \in [0,\beta]$ with the inverse temperature $\beta=1/T$.
In the Wick rotation, $\partial_{t}$ is replaced to $\partial_{\tau}=\partial/\partial \tau$ whose relation is $\partial_{t}=i\partial_{\tau}$.
The generating functional is given by
\begin{eqnarray}
   \bar{Z}
&=&
  \int {\cal D}\varphi {\cal D}\varphi^{\dag}
  \exp
  \Biggl[
      -
       \int \drm\tau \, \drm^{3}\vec{x} \,
       \Biggl(
              \varphi^{\dag}(\tau,\vec{x})
              \biggl(
                       \partial_{\tau}
                    - \frac{\vec{\nabla}^{2}}{2m}
                    - \mu
                    - \vec{\mu}_{n} \!\cdot\! \vec{B}
              \biggr)
              \varphi(\tau,\vec{x})
           - G \sum_{a,b} T^{ab\dag}(\tau,\vec{x})T^{ab}(\tau,\vec{x})
       \Biggr)
  \Biggr],
\end{eqnarray}
in the path-integral formalism, where ${\cal D}\varphi^{\ast}$ and ${\cal D}\varphi$ indicate the path integrals for $\varphi^{\ast}$ and $\varphi$, respectively.
The calculation of the path-integral is difficult to be performed in an analytic form due to the interaction term, and hence an approximate solution should be searched.
For this purpose, we introduce a complex tensor field $A^{ab}(\tau,\vec{x})$ ($a,b=1,2,3$) as an auxiliary field, where
$A^{ab}(\tau,\vec{x})$ is a symmetric and traceless tensor.
It will turn out that $A^{ab}(\tau,\vec{x})$ is the order parameter relevant to the neutron pairing induced by the tensor term in Eq.~\eqref{eq:tensor_def}.
The inclusion of the auxiliary field into the Lagrangian is accomplished as follows.
First, we multiply
\begin{eqnarray}
 \int {\cal D} A {\cal D} A^{\dag}
 \exp
 \Biggl(
     - \frac{1}{G}
       \int \drm\tau \, \drm^{3}\vec{x} \, \sum_{a,b} A^{ab\dag}(\tau,\vec{x}) A^{ab}(\tau,\vec{x})
 \Biggr)
\label{eq:auxiliary_field_introduction}
\end{eqnarray}
to the generating functional $\bar{Z}$,
and shift $A^{ab}(\tau,\vec{x})$ by
\begin{eqnarray}
 A^{ab}(\tau,\vec{x})
\rightarrow
 A^{ab}(\tau,\vec{x}) + G \, T^{ab}(\tau,\vec{x}).
\end{eqnarray}
Notice that the quantity in Eq.~\eqref{eq:auxiliary_field_introduction} is a constant number, because the path-integral is performed for $A^{ab}$
and $A^{ab\dag}$ as it is represented by ${\cal D}A$ and ${\cal D}A^{\dag}$, respectively.
Thus, the multiplication of Eq.~\eqref{eq:auxiliary_field_introduction} to the generating functional does not change the contents of physics.
Notice also that the above shift of $A^{ab}(\tau,\vec{x})$ is always possible due to the arbitrariness of parametrizations of the auxiliary field.
In the end, we obtain a new form of the generating functional
\begin{eqnarray}
  Z
&=&
  \int {\cal D}\varphi {\cal D}\varphi^{\dag} {\cal D}A {\cal D}A^{\dag}
  \exp
  \Biggl[
      -
       \Biggl(
              \int \drm\tau \, \drm^{3}\vec{x} \,
              \varphi^{\dag}(\tau,\vec{x})
              \biggl( \partial_{\tau} - \frac{\vec{\nabla}^{2}}{2m} - \mu - \vec{\mu}_{n} \!\cdot\! \vec{B} \biggr)
              \varphi(\tau,\vec{x})
              \nonumber \\ && \hspace{10em}
           + \sum_{a,b} A^{ab\dag}(\tau,\vec{x}) T^{ab}(\tau,\vec{x})
           + \sum_{a,b} T^{ab\dag}(\tau,\vec{x})  A^{ab}(\tau,\vec{x})
           + \frac{1}{G}
              \sum_{a,b} A^{ab\dag}(\tau,\vec{x}) A^{ab}(\tau,\vec{x})
       \Biggr)
  \Biggr],
\label{eq:generating_functional_aux_3m}
\end{eqnarray}
where the neutron field $\varphi$ and the auxiliary field $A^{ab}$ are coupled to each other.
$Z$ is physically the same as the original generating functional $\bar{Z}$.
In the next subsection, we will integrate out the neutron field and construct the effective Lagrangian containing only the auxiliary field at the low-energy scale.

\subsection{Nambu-Gor'kov representation}

In performing the path-integral for the neutron field $\varphi$ in Eq.~\eqref{eq:generating_functional_aux_3m},
it is convenient to represent $\varphi$ in the Nambu-Gor'kov representation.
For this purpose, we define a four-component spinor field
\begin{eqnarray}
 \Psi(\tau,\vec{x})
=
\frac{1}{\sqrt{2}}
\left(
\begin{array}{c}
 \varphi(\tau,\vec{x}) \\
 \varphi^{\ast}(\tau,\vec{x}) 
\end{array}
\right),
\label{eq:field_NG_formalism}
\end{eqnarray}
which contains the particle component $\varphi$ in the first and second columns and 
the hole component $\varphi^{\ast}$ in the third and fourth columns.
With this field, the generating function can be expressed by
\begin{eqnarray}
  Z
=
  \int {\cal D}\Psi {\cal D}\Psi^{\dag} {\cal D}A {\cal D}A^{\dag}
  \exp
  \Biggl[
      - \int \drm\tau \, \drm^{3}\vec{x} \,
        \biggl(
              \Psi^{\dag}(\tau,\vec{x}) S(\tau,\vec{x})^{-1} \Psi(\tau,\vec{x})
          + \frac{1}{4G} \sum_{a,b} A^{ab\dag}(\tau,\vec{x}) A^{ab}(\tau,\vec{x})
        \biggr)
  \Biggr].
\label{eq:generating_functional_Z0}
\end{eqnarray}
Here, we have defined the inverse of the propagator in the Nambu-Gor'kov space by
\begin{eqnarray}
  S(\tau,\vec{x})^{-1}
\equiv
   \left(
    \begin{array}{cc}
          \partial_{\tau} - \dfrac{\vec{\nabla}^{2}}{2m} - \mu - \vec{\mu}_{n} \!\cdot\! \vec{B}
          & - {\displaystyle \sum_{a,b}} A^{ab}(\tau,\vec{x}) t^{ab\dag}(\vec{x}) \\
          {\displaystyle \sum_{a,b}} A^{ab\dag}(\tau,\vec{x}) t^{ab}(\vec{x})
          & \partial_{\tau} + \dfrac{\vec{\nabla}^{2}}{2m} + \mu + \vec{\mu}_{n}^{t} \!\cdot\! \vec{B}
    \end{array}
    \right),
\label{eq:S_propagator_def}
\end{eqnarray}
where we have introduced
\begin{eqnarray}
  t^{ab}(\vec{x})
\equiv
- \biggl( \frac{1}{2}\Sigma^{a\dag} \nabla^{b} + \frac{1}{2}\Sigma^{b\dag} \nabla^{a} - \frac{1}{3}\delta^{ab}\sum_{c}\Sigma^{c\dag} \nabla^{c} \biggr),
\label{eq:t_def}
\end{eqnarray}
for the coupling of the spin and the momentum.
It is convenient to notice $\vec{\sigma}^{t}=-\sigma^{2}\vec{\sigma}\sigma^{2}$ in the following calculation.
The first term in the exponential in Eq.~\eqref{eq:generating_functional_Z0} is a bilinear form for $\Psi$ and $\Psi^{\dag}$, and hence it is possible to perform the integral analytically.
The result is
\begin{eqnarray}
  \int {\cal D}\Psi {\cal D}\Psi^{\dag}
  \exp
  \Biggl[
      - \int \drm\tau \, \drm^{3}\vec{x} \,
        \biggl(
              \Psi^{\dag}(\tau,\vec{x}) S(\tau,\vec{x})^{-1} \Psi(\tau,\vec{x})
        \biggr)
  \Biggr]
=
 \mathrm{Det} \, \bigl( S(\tau,\vec{x})^{-1} \bigr),
\end{eqnarray}
where $\mathrm{Det}$ indicates the determinant of the matrix whose indices are given not only by the spin of the neutron but also by the imaginary time $\tau$ and the three-dimensional space $\vec{x}$. 
Furthermore, adopting the formula $\mathrm{Det} \, M=e^{\mathrm{Tr} \ln M}$ which is applicable to a matrix $M$,
we can express the generating functional \eqref{eq:generating_functional_Z0} as
\begin{eqnarray}
  Z
=
  \int {\cal D}A {\cal D}A^{\dag}
  \exp
  \Biggl[ 
        \mathrm{Tr} \ln \bigl( S(\tau,\vec{x})^{-1} \bigr)
      - \frac{1}{4G}\int \drm\tau \, \drm^{3}\vec{x} \,
        \sum_{a,b}
        A^{ab\dag}(\tau,\vec{x}) A^{ab}(\tau,\vec{x})
  \Biggr],
\label{eq:generating_functional_Z}
\end{eqnarray}
where only the path-integral for $A$ and $A^{\dag}$ is left.
However, the above form of the generating functional does not provide directly an answer to the question about the ground state of the system.
This is because the first term in the exponential $\mathrm{Tr} \ln \bigl( S(\tau,\vec{x})^{-1} \bigr)$ contains the field $A(\tau,\vec{x})$ and $A^{\dag}(\tau,\vec{x})$, and hence it is still difficult to perform the path-integral for $A$ and $A^{\dag}$.
Instead of the complete calculation,
we will expand $\mathrm{Tr} \ln \bigl( S(\tau,\vec{x})^{-1} \bigr)$ for a small $A(\tau,\vec{x})$  and $A^{\dag}(\tau,\vec{x})$, and will try to obtain approximately an analytical equation applicable for the small $A(\tau,\vec{x})$ around the critical temperature.
We also expand $\mathrm{Tr} \ln \bigl( S(\tau,\vec{x})^{-1} \bigr)$ in terms of a small magnetic field $\vec{B}$.
In the previous work by ones of the authors, the higher-order terms of the magnetic field was investigated, and the influence of the strong magnetic field on the phase diagram was studied~\cite{Yasui:2018tcr,Yasui:2019tgc}.
The present calculation is the extension to the 8th order of $A(\tau,\vec{x})$.

\subsection{Effective potential}

For the generating function $Z$ in Eq.~\eqref{eq:generating_functional_Z}, we define the effective potential $W$ by $Z=e^{-W}$ where only the one-particle irreducible diagrams are contained and the other diagrams are removed.
By regarding $A(\tau,\vec{x})$ as a classical field (mean-field) and comparing $W$ with the functions in the exponential in Eq.~\eqref{eq:generating_functional_Z}, we obtain the effective potential
\begin{eqnarray}
 W
=
 - \mathrm{Tr} \ln \bigl( S(\tau,\vec{x})^{-1} \bigr)
+ \frac{1}{4G} \int \drm\tau \, \drm^{3}\vec{x} \,
   \sum_{a,b} A^{ab\dag}(\tau,\vec{x}) A^{ab}(\tau,\vec{x}).
\end{eqnarray}
Notice that the mean-field approximation was adopted, where the fluctuations about $A(\tau,\vec{x})$ are not considered and hence the strong-coupling effect is not included.
In order to expand the trace in terms of $A(\tau,\vec{x})$ and the magnetic field $\vec{B}$,
we decompose the inverse of the propagator as $\displaystyle S(\tau,\vec{x})^{-1} =  S_{m}(\tau,\vec{x})^{-1} + \hat{A}(\tau,\vec{x})$.
Here $S_{m}(\tau,\vec{x})^{-1}$ is the part including the free term and the spin-magnetic interaction term
\begin{eqnarray}
   S_{m}(\tau,\vec{x})^{-1}
=
    \left(
    \begin{array}{cc}
        -i\omega_{N} + \xi_{\vec{p}} - \vec{\mu}_{n} \!\cdot\! \vec{B} & 0 \\
        0 & -i\omega_{N} - \xi_{\vec{p}} + \mu + \vec{\mu}_{n}^{t} \!\cdot\! \vec{B}
    \end{array}
    \right),
\end{eqnarray}
and $ \hat{A}(\tau,\vec{x})$ is the part which contains the auxiliary field $A^{ab}(\tau,\vec{x})$
\begin{eqnarray}
   \hat{A}(\tau,\vec{x})
&=&
   \left(
    \begin{array}{cc}
          0 & -{\displaystyle \sum_{a,b}}A^{ab}(\tau,\vec{x}) t^{ab\dag}(\vec{x}) \\
          {\displaystyle \sum_{a,b}}A^{ab\dag}(\tau,\vec{x}) t^{ab}(\vec{x}) & 0
    \end{array}
    \right).
\label{eq:tau_matrix_def}
\end{eqnarray}
By adopting the expansion for the logarithmic function,
we finally obtain an expanded form of the effective potential
\begin{eqnarray}
 W
=
  - \mathrm{Tr} \ln \bigl( S_{m}(\tau,\vec{x})^{-1} \bigr)
  - \sum_{N\ge1} \frac{(-1)^{N+1}}{N} \mathrm{Tr} \Bigl( \bigl( S_{m}(\tau,\vec{x}) \hat{A}(\tau,\vec{x}) \bigr)^{N} \Bigr)
 + \frac{1}{4G} \int \drm\tau \, \drm^{3}\vec{x} \,
    \sum_{a,b} A^{ab\dag}(\tau,\vec{x}) A^{ab}(\tau,\vec{x}).
\label{eq:eff_pot_weakcoupling_b}
\end{eqnarray}
We notice that trace, $\mathrm{Tr}$, takes the sum over all the variables such as spin, imaginary time and space.
The concrete procedure of the trace calculation is presented in Appendix~\ref{sec:trace_calculation}.
For the later convenience, we define the density of the effective potential $w[A(\tau,\vec{x})]$ by
\begin{eqnarray}
   W = \int \drm \tau \, \drm^{3}\vec{x} \, w[A(\tau,\vec{x})].
\label{eq:eff_pot_density}
\end{eqnarray}
This quantity will be used in the next subsection.

\subsection{Ginzburg-Landau free energy}
\label{sec:GL_free_energy}

By using the density of the effective potential $w[A(\tau,\vec{x})]$ in Eq.~\eqref{eq:eff_pot_density}, we define the GL free energy $f[A(\tau,\vec{x})]$ by
\begin{eqnarray}
   f[A(\tau,\vec{x})] = \frac{1}{2} w[A(\tau,\vec{x})],
\end{eqnarray}
where the coefficient $1/2$ is multiplied in order to avoid the double-counting of the fermionic degrees of freedom in the Nambu-Gor'kov representation (cf.~Eq.~\eqref{eq:field_NG_formalism}).
By adopting the expansion for the boson field $A^{ab}$ and the magnetic field $\vec{B}$ in Eq.~\eqref{eq:eff_pot_weakcoupling_b},
we obtain the free energy, given by
\begin{eqnarray}
  f[{A}] = f_{0} + f_{8}^{(0)}[{A}] + f_{2}^{(\le4)}[{A}] + f_{4}^{(\le2)}[{A}] + {\cal O}(B^{m}{A}^{n})_{m+n\ge7}.
\label{eq:eff_pot_coefficient02_f}
\end{eqnarray}
as a series of $A$ and $\vec{B}$.
Each function is defined as follows.
$f_{0}$ is the sum of the free part and the spin-magnetic coupling term
\begin{eqnarray}
f_{0}
=
  - T
    \int \frac{\mathrm{d}^{3}\vec{p}}{(2\pi)^{3}}
    \ln \Bigl(
             \bigl( 1+e^{-\xi_{\vec{p}}^{-}/T} \bigr)
             \bigl( 1+e^{-\xi_{\vec{p}}^{+}/T} \bigr)
         \Bigr),
\label{eq:eff_pot_free_magneticfield}
\end{eqnarray}
with $\displaystyle \xi_{\vec{p}}^{\pm} = \xi_{\vec{p}} \pm |\vec{\mu}_{n}||\vec{B}|$ and $\displaystyle \xi_{\vec{p}}=\vec{p}^{2}/(2m)-\mu$.
The other terms include the condensate $A^{ab}$.
$f_{8}^{(0)}[{A}]$ consists of the terms including the field $A^{ab}$ up to the 8th order with no magnetic field,
$f_{2}^{(\le4)}[{A}]$ consists of the terms including the field $A^{ab}$ up to the 2nd order with the magnetic field up to $|\vec{B}|^{4}$, 
and $f_{4}^{(\le2)}[{A}]$ consists of the terms including the field $A^{ab}$ up to the 4th order with the magnetic field up to $|\vec{B}|^{2}$.
Their explicit forms are given by
\begin{eqnarray}
 f_{8}^{(0)}[{A}]
&=&
   \alpha^{(0)}
   \bigl(\mathrm{tr} {A}^{\ast} {A} \bigr)
\nonumber \\ &&
+ K^{(0)}
  \Bigl(
        \nabla_{xi} {A}^{ba\ast}
        \nabla_{xi} {A}^{ab}
     + \nabla_{xi} {A}^{ia\ast}
        \nabla_{xj} {A}^{aj}
     + \nabla_{xi} {A}^{ja\ast}
        \nabla_{xj} {A}^{ai}
  \Bigr)
\nonumber \\ &&
+ \beta^{(0)}
   \Bigl(
        \bigl(\mathrm{tr} \, {A}^{\ast} {A} \bigr)^{2}
      - \bigl(\mathrm{tr} \, {A}^{\ast 2} {A}^{2} \bigr)
   \Bigr)
\nonumber \\ &&
+ \gamma^{(0)}
   \Bigl(
         - 3  \bigl(\mathrm{tr} \, {A}^{\ast} {A} \bigr) \bigl(\mathrm{tr} \, {A}^{2} \bigr) \bigl(\mathrm{tr} \, {A}^{\ast 2} \bigr)
        + 4 \bigl(\mathrm{tr} \, {A}^{\ast} {A} \bigr)^{3}
        + 6 \bigl(\mathrm{tr} \, {A}^{\ast} {A} \bigr) \bigl(\mathrm{tr} \, {A}^{\ast 2} {A}^{2} \bigr)
      + 12 \bigl(\mathrm{tr} \, {A}^{\ast} {A} \bigr) \bigl(\mathrm{tr} \, {A}^{\ast} {A} {A}^{\ast} {A} \bigr)
              \nonumber \\ && \hspace{3em} 
         - 6 \bigl(\mathrm{tr} \, {A}^{\ast 2} \bigr) \bigl(\mathrm{tr} \, {A}^{\ast} {A}^{3} \bigr)
         - 6 \bigl(\mathrm{tr} \, {A}^{2} \bigr) \bigl(\mathrm{tr} \, {A}^{\ast 3} {A} \bigr)
       - 12 \bigl(\mathrm{tr} \, {A}^{\ast 3} {A}^{3} \bigr)
      + 12 \bigl(\mathrm{tr} \, {A}^{\ast 2} {A}^{2} {A}^{\ast} {A} \bigr)
        + 8 \bigl(\mathrm{tr} \, {A}^{\ast} {A} {A}^{\ast} {A} {A}^{\ast} {A} \bigr)
   \Bigr)
\nonumber \\ &&
 + \delta^{(0)}
\Bigl(
       \bigl( \mathrm{tr}\,A^{\ast 2} \bigr)^{2} \bigl( \mathrm{tr}\, A^{2} \bigr)^{2}
 + 2 \bigl( \mathrm{tr}\,A^{\ast 2} \bigr)^{2} \bigl( \mathrm{tr}\, A^{4} \bigr)
  - 8 \bigl( \mathrm{tr}\,A^{\ast 2} \bigr) \bigl( \mathrm{tr}\,A^{\ast}AA^{\ast}A \bigr) \bigl( \mathrm{tr}\,A^{2} \bigr)
  - 8 \bigl( \mathrm{tr}\,A^{\ast 2} \bigr) \bigl( \mathrm{tr}\,A^{\ast}A \bigr)^{2} \bigl( \mathrm{tr}\,A^{2} \bigr)
       \nonumber \\ && \hspace{3em}
 - 32 \bigl( \mathrm{tr}\,A^{\ast 2} \bigr) \bigl( \mathrm{tr}\,A^{\ast}A \bigr) \bigl( \mathrm{tr}\,A^{\ast}A^{3} \bigr)
 - 32 \bigl( \mathrm{tr}\,A^{\ast 2} \bigr) \bigl( \mathrm{tr}\,A^{\ast}AA^{\ast}A^{3} \bigr)
 - 16 \bigl( \mathrm{tr}\,A^{\ast 2} \bigr) \bigl( \mathrm{tr}\,A^{\ast}A^{2}A^{\ast}A^{2} \bigr)
       \nonumber \\ && \hspace{3em}
  + 2 \bigl( \mathrm{tr}\,A^{\ast 4} \bigr) \bigl( \mathrm{tr}\,A^{2} \bigr)^{2}
  + 4 \bigl( \mathrm{tr}\,A^{\ast 4} \bigr) \bigl( \mathrm{tr}\,A^{4} \bigr)
  - 32 \bigl( \mathrm{tr}\,A^{\ast 3}A \bigr) \bigl( \mathrm{tr}\,A^{\ast}A \bigr) \bigl( \mathrm{tr}\,A^{2} \bigr)
       \nonumber \\ && \hspace{3em}
  - 64 \bigl( \mathrm{tr}\,A^{\ast 3}A \bigr) \bigl( \mathrm{tr}\,A^{\ast}A^{3} \bigr)
  - 32 \bigl( \mathrm{tr}\,A^{\ast 3}AA^{\ast}A \bigr) \bigl( \mathrm{tr}\,A^{2} \bigr)
  - 64 \bigl( \mathrm{tr}\,A^{\ast 3}A^{2}A^{\ast}A^{2} \bigr)
  - 64 \bigl( \mathrm{tr}\,A^{\ast 3}A^{3} \bigr) \bigl( \mathrm{tr}\,A^{\ast}A \bigr)
       \nonumber \\ && \hspace{3em}
  - 64 \bigl( \mathrm{tr}\,A^{\ast 2}AA^{\ast 2}A^{3} \bigr)
  - 64 \bigl( \mathrm{tr}\,A^{\ast 2}AA^{\ast}A^{2} \bigr) \bigl( \mathrm{tr}\,A^{\ast}A \bigr)
 + 16 \bigl( \mathrm{tr}\,A^{\ast 2}A^{2} \bigr)^{2}
 + 32 \bigl( \mathrm{tr}\,A^{\ast 2}A^{2} \bigr) \bigl( \mathrm{tr}\,A^{\ast}A \bigr)^{2}
       \nonumber \\ && \hspace{3em}
 + 32 \bigl( \mathrm{tr}\,A^{\ast 2}A^{2} \bigr) \bigl( \mathrm{tr}\,A^{\ast}AA^{\ast}A \bigr)
 + 64 \bigl( \mathrm{tr}\,A^{\ast 2}A^{2}A^{\ast 2}A^{2} \bigr)
  -16 \bigl( \mathrm{tr}\,A^{\ast 2}AA^{\ast 2}A \bigr) \bigl( \mathrm{tr}\,A^{2} \bigr)
   + 8 \bigl( \mathrm{tr}\,A^{\ast}A \bigr)^{4}
       \nonumber \\ && \hspace{3em}
 + 48 \bigl( \mathrm{tr}\,A^{\ast}A \bigr)^{2} \bigl( \mathrm{tr}\,A^{\ast}AA^{\ast}A \bigr)
 +192 \bigl( \mathrm{tr}\,A^{\ast}A \bigr) \bigl( \mathrm{tr}\,A^{\ast}AA^{\ast 2}A^{2} \bigr)
 + 64 \bigl( \mathrm{tr}\,A^{\ast}A \bigr) \bigl( \mathrm{tr}\,A^{\ast}AA^{\ast}AA^{\ast}A \bigr)
       \nonumber \\ && \hspace{3em}
  -128 \bigl( \mathrm{tr}\,A^{\ast}AA^{\ast 3}A^{3} \bigr)
 + 64 \bigl( \mathrm{tr}\,A^{\ast}AA^{\ast 2}AA^{\ast}A^{2} \bigr)
 + 24 \bigl( \mathrm{tr}\,A^{\ast}AA^{\ast}A \bigr)^{2}
 +128 \bigl( \mathrm{tr}\,A^{\ast}AA^{\ast}AA^{\ast 2}A^{2} \bigr)
       \nonumber \\ && \hspace{3em}
 + 48 \bigl( \mathrm{tr}\,A^{\ast}AA^{\ast}AA^{\ast}AA^{\ast}A \bigr)
\Bigr),
\label{eq:eff_pot_w0_coefficient02_f}
\\
   f_{2}^{(\le4)}[{A}]
&=&
      \beta^{(2)}
      \vec{B}^{t} {A}^{\ast} {A} \vec{B}
+ \beta^{(4)}
   |\vec{B}|^{2}
   \vec{B}^{t} {A}^{\ast} {A} \vec{B},
\label{eq:eff_pot_B4w2_coefficient02_f}
\\
   f_{4}^{(\le2)}[{A}]
&=&
  \gamma^{(2)}
  \Bigl(
       - 2 \, |\vec{B}|^{2} \bigl(\mathrm{tr} \, {A}^{2} \bigr) \bigl(\mathrm{tr} \, {A}^{\ast 2} \bigr)
       - 4 \, |\vec{B}|^{2} \bigl(\mathrm{tr} \, {A}^{\ast} {A} \bigr)^{2}
      + 4 \, |\vec{B}|^{2} \bigl(\mathrm{tr} \, {A}^{\ast} {A} {A}^{\ast} {A} \bigr)
      + 8 \, |\vec{B}|^{2} \bigl(\mathrm{tr} \, {A}^{\ast 2} {A}^{2} \bigr)
            \nonumber \\ && \hspace{2em}
        + \vec{B}^{t} {A}^{2} \vec{B} \bigl(\mathrm{tr} \, {A}^{\ast 2} \bigr)
       - 8 \, \vec{B}^{t} {A}^{\ast} {A} \vec{B} \bigl(\mathrm{tr} \, {A}^{\ast} {A} \bigr)
         + \vec{B}^{t} {A}^{\ast 2} \vec{B} \bigl(\mathrm{tr} \, {A}^{2} \bigr)
      + 2 \, \vec{B}^{t} {A} {A}^{\ast 2} {A} \vec{B}
            \nonumber \\ && \hspace{2em}
      + 2 \, \vec{B}^{t} {A}^{\ast} {A}^{2} {A}^{\ast} \vec{B}
       - 8 \, \vec{B}^{t} {A}^{\ast} {A} {A}^{\ast} {A} \vec{B}
       - 8 \, \vec{B}^{t} {A}^{\ast 2} {A}^{2} \vec{B}
  \Bigr).
\label{eq:eff_pot_B2w4_coefficient02_f}
\end{eqnarray}
Here, the GL coefficients can be obtained as
\begin{eqnarray}
   \alpha^{(0)}
&=&
   \frac{N(0)p_{F}^{2}}{3} \log\frac{T}{T_{c0}},
\nonumber \\ 
  K^{(0)}
&=&
   \frac{7 \, \zeta(3)N(0) p_{F}^{4}}{240m^{2}(\pi T)^{2}},
\nonumber \\ 
  \beta^{(0)}
&=&
   \frac{7\,\zeta(3)N(0)p_{F}^{4}}{60\,(\pi T)^{2}},
\nonumber \\ 
  \gamma^{(0)}
&=&
   - \frac{31\,\zeta(5)N(0)p_{F}^{6}}{13440\,(\pi T)^{4}},
\nonumber \\ 
  \delta^{(0)}
&=&
  \frac{127\,\zeta(7)N(0)p_{F}^{8}}{387072\,(\pi T)^{6}},
\nonumber \\ 
   \beta^{(2)}
&=&
    \frac{7\,\zeta(3)N(0)p_{F}^{2}\gamma_{n}^{2}}{48(1+F_{0}^{a})^{2}(\pi T)^{2}},
\nonumber \\ 
   \beta^{(4)}
&=&
    - \frac{31\,\zeta(5)N(0)p_{F}^{2}\gamma_{n}^{4}}{768(1+F_{0}^{a})^{4}(\pi T)^{4}},
\nonumber \\ 
  \gamma^{(2)}
&=&
  \frac{31\,\zeta(5)N(0)p_{F}^{4}\gamma_{n}^{2}}{3840(1+F_{0}^{a})^{2}(\pi T)^{4}},
\label{eq:eff_pot_coefficient0_parameters_FL_f}
\end{eqnarray}
with $N(0)=m\,p_{F}/(2\pi^{2})$ the state-number-density at the Fermi surface and $|\vec{\mu}_{n}^{\ast}|=(\gamma_{n}/2)/(1+F_{0}^{a})$ the magnitude of the magnetic momentum of a neutron modified by the Landau parameter $F_{0}^{a}$.
The Landau parameter stems from the Hartree-Fock approximation which are not considered explicitly in the present formalism.
$\zeta(n)$ is the zeta function.
We use the condition that the temperature $T$ is close to the critical temperature at zero magnetic field $T_{c0}$; $|1-T/T_{c0}| \ll 1$.
The critical temperature is related to the strength of the coupling constants through
\begin{eqnarray}
 T_{c0}
=
 \frac{\pi e^{-\gamma}D}{8}
 \exp \biggl( -\frac{3}{8N(0)p_{F}^{2}G} \biggr),
\end{eqnarray}
with the Euler's constant $\gamma$, where the energy scale $D$ relevant for the typical scales in the dynamics is introduced as the cutoff parameter in the momentum space below and above the Fermi surface.
We notice that the parameters of the coupling constant $G$ and the momentum cutoff $D$ are controlled by the critical temperature $T_{c0}$ through the above relation.
Therefore, $T_{c0}$ is the only parameter to control the energy scale of the neutron $^{3}P_{2}$ superfluidity.
We remind us that this simplification is just due to the one-loop approximation for the neutron part in the calculation of the GL free energy density.
In the derivation of the above equations, we have used the quasi-classical approximation for the momentum integrals.
The concrete calculations are shown in Appendix~\ref{sec:trace_calculation}.
We emphasize that the 8th order terms, {\it i.e.}, the terms with the coefficient $\delta^{(0)}$, 
provide a new ingredient which has not been calculated so far in the literature.

\section{Stability of the ground state}
\label{sec:stability}

\begin{figure}[tb]
\begin{center}
\includegraphics[scale=0.3]{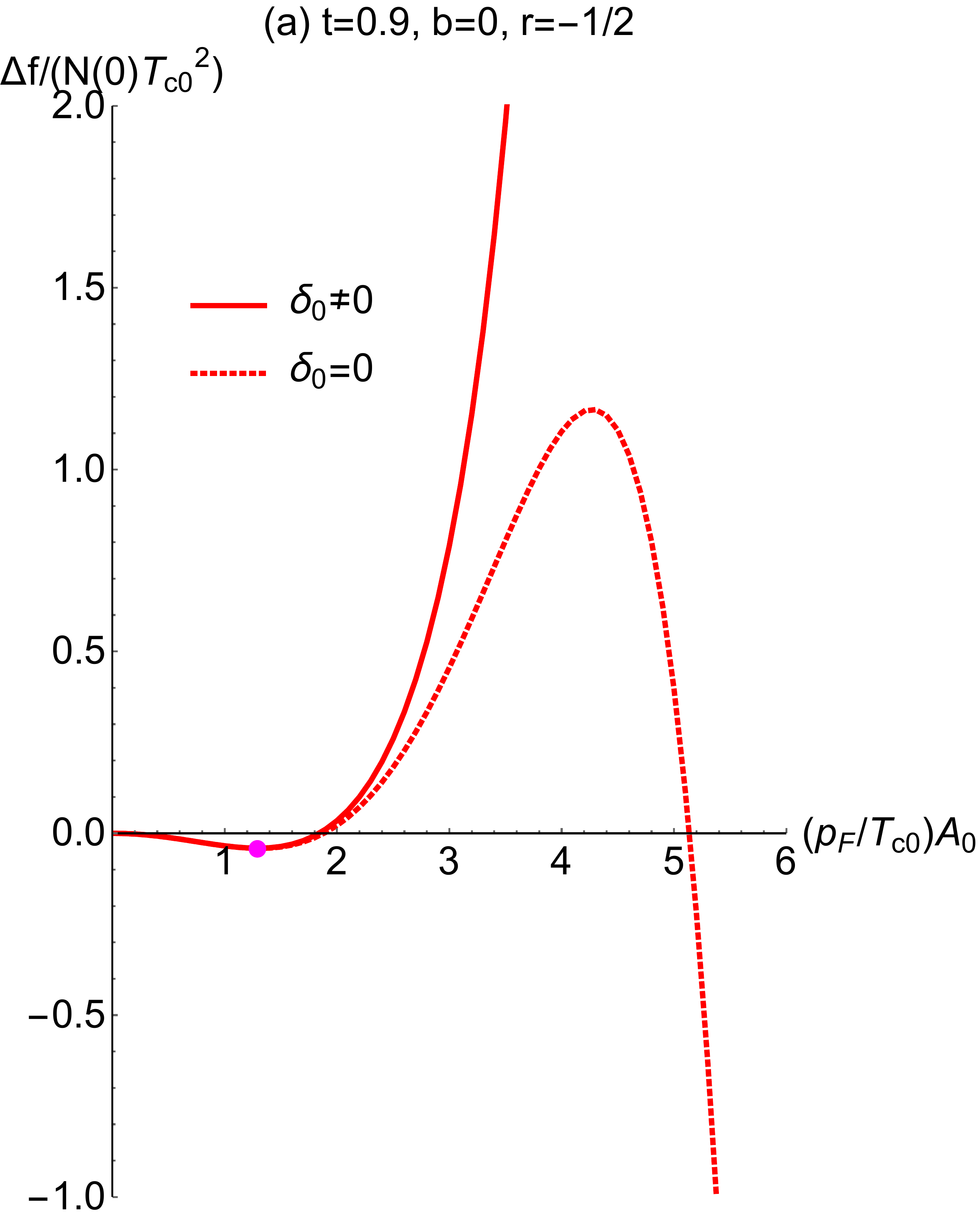}
\hspace{1em}
\includegraphics[scale=0.3]{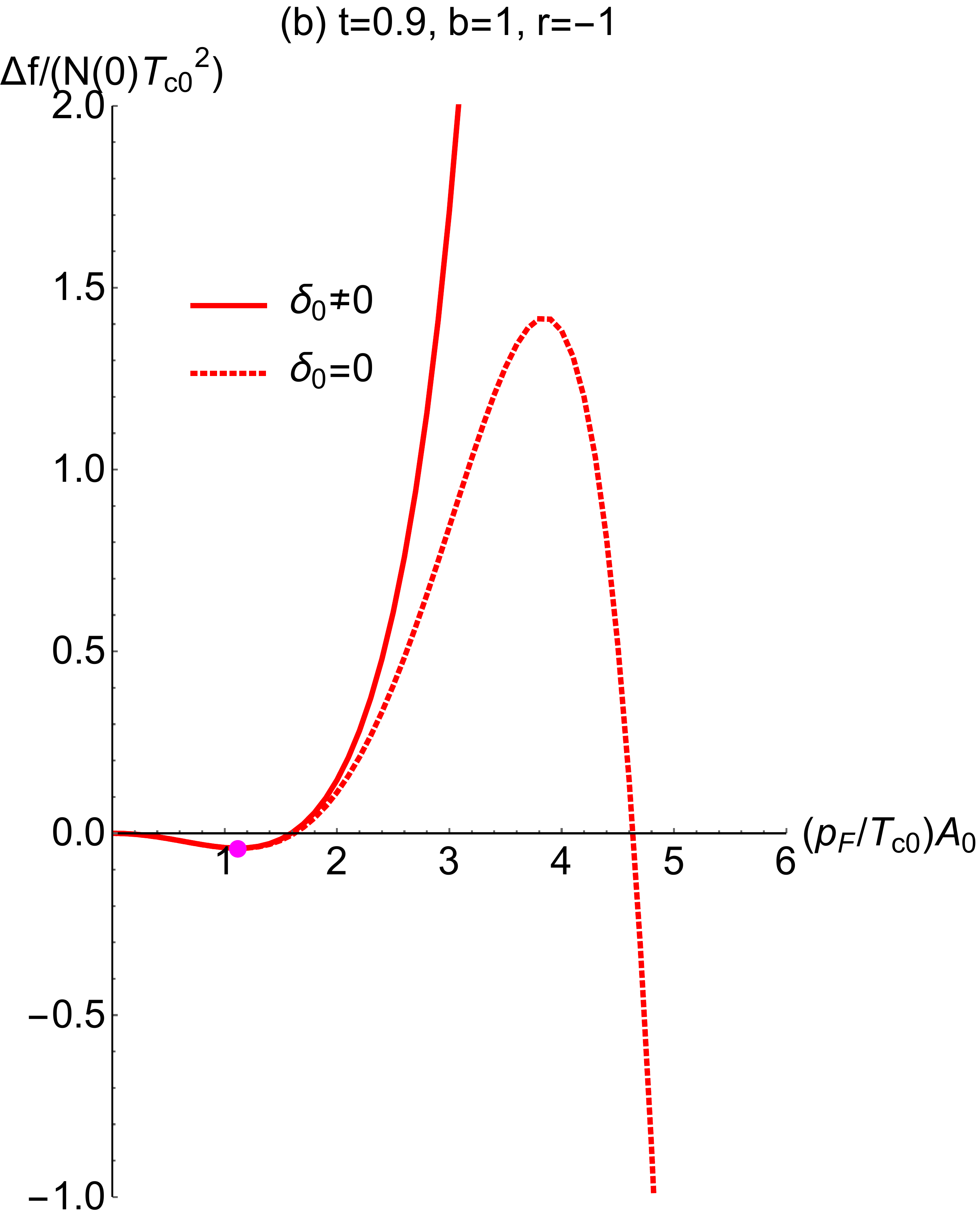}
\caption{The plots of the GL free energy density $\Delta f=f-f_{0}$ as functions of $A_{0}$ for $t=0.9$. (a) $b=0$ ($r=-1/2$) and (b) $b=1$ ($r=-1$).  The dotted lines are the results in the previous study (setting $\delta^{(0)}=0$ in Eq.~(\ref{eq:eff_pot_coefficient02_f})), and the solid lines are the results up to ${\cal O}({A}^{8})$ in the present study (setting $\delta^{(0)}\neq0$). The magenta points are the minimum points with respect to $A_{0}$.}
\label{fig:Fig_f_b0}
\end{center}
\end{figure}

In this section, as one of the most important applications of the 8th order terms 
that we have obtained in the last section, 
we show  
that the GL expansion up to the 8th order admits the stable and unique ground state 
once the temperature and magnetic field are fixed, 
while the expansion up to 6th order has the instability.
We plot the GL free energy density, $\Delta f=f-f_{0}$ (the difference between the full equation of the GL free energy density and the part from no condensate terms), as a function of $A_{0}$ for fixed $t$, $b$, and $r$ in Fig.~\ref{fig:Fig_f_b0}.
We exemplify the two cases: (i) $t=0.9$, $b=0$, and $r=-1/2$, and (ii) $t=0.9$, $b=1$, and $r=-1$.
The former denotes the UN phase, and the latter the D$_{4}$-BN phase.
The GL free energy function with $\delta_{0}=0$ is presented by the dashed lines in the figure.
The (locally) minimum points of the GL free energy density are presented by the magenta blobs.
As for the location of the minimum points, there is no much difference between the $\delta_{0}=0$ and $\delta_{0}\neq0$ cases, as the behaviors for the smaller $A_{0}$ are observed to be almost the same for the both cases.
However, we notice that the behaviors at larger $A_{0}$ are different to each other:
$\Delta f$ has the instability in the $\delta_{0}=0$ case, and $\Delta f$ has no such instability in the $\delta_{0}\neq0$ case.
The former leads to the stability of the ground state as the local minimum, while the latter to the global minimum.
The existence of the global minimum is important to ensure the stability of the ground state.
It is also practically useful to search the minimal points in the variational calculations in the wide range of the parameters.
Thus, the GL expansion up the 8th order is necessary for the system having the stable and unique ground state, 
and it helps
 us to investigate easily various properties of the neutron $^{3}P_{2}$ superfluidity.

\section{Phase diagram}
\label{sec:phase_diagram}

With the GL free energy density \eqref{eq:eff_pot_coefficient02_f},
we show the phase diagram of the neutron $^{3}P_{2}$ superfluidity on the plane spanned by the temperature and the magnetic field.
We consider the nematic phases in which the order parameter $A^{ab}$ is parametrized by
\begin{eqnarray}
  A^{ab}
=
A_{0}
\left(
\begin{array}{ccc}
 r & 0  & 0  \\
 0 & -1-r  & 0  \\
 0 & 0  & 1  
\end{array}
\right)_{ab},
\label{eq:A_gs}
\end{eqnarray}
with $A_{0}\ge0$ being the magnitude of the condensate and $r$ is 
a real parameter 
which can be restricted to $-1\le r \le -1/2$ by using the $SO(3) \times U(1)$ symmetry without the loss of generality.
We notice that it is always possible to transform any symmetric and traceless matrix into the diagonal form in which the diagonal elements are aligned in the order of the magnitudes of the absolute values. 
For the different values of $r$, the order parameter possesses different symmetries:
the uniaxial-nematic (UN) phase ($r=-1/2$) and the biaxial-nematic (BN) phase ($-1\le r < -1/2$).
The UN phase has the U(1) symmetry.
The BN phase is further classified to the D$_{2}$-BN phase ($-1<r<-1/2$) and the D$_{4}$-BN phase ($r=-1$), 
having the D$_{2}$ and D$_{4}$ dihedral symmetries, respectively.
The explicit forms of group transformations of those symmetries are summarized in Appendix~\ref{sec:symmetries}.
We substitute Eq.~\eqref{eq:A_gs} to the GL free energy density \eqref{eq:eff_pot_coefficient02_f}, and perform the variational calculation with respect to $A_{0}$ and $r$.
In the present study, we choose the direction of the magnetic field along the $y$-axis, $\vec{B}=(0,B,0)$, to minimize the total energy.

\begin{figure}[tb]
\begin{center}
\includegraphics[scale=0.5]{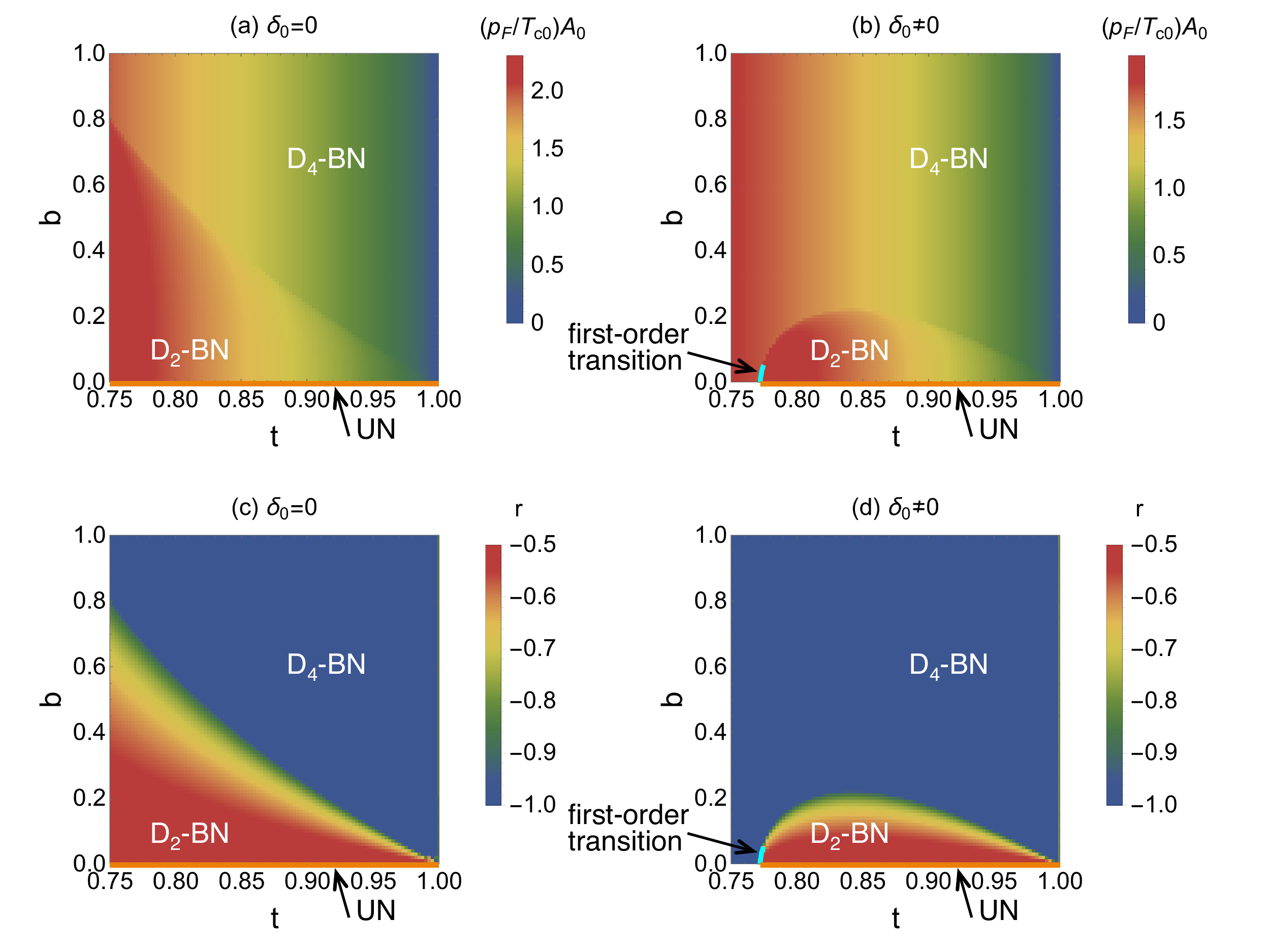}
\caption{The phase diagram for ${A}_{0}$ and $r$ on the plane spanned by the temperature ($t=T/T_{c0}$) and the magnetic field ($b=\gamma_{n}B/((1+F_{0}^{a})T_{c0})$). The two left panels (a, c) are the results in the previous study (setting $\delta^{(0)}=0$ in Eq.~(\ref{eq:eff_pot_coefficient02_f}))~\cite{Yasui:2018tcr,Yasui:2019tgc}, and the two right panels (b, d) are the results up to ${\cal O}({A}^{8})$ in the present study (setting $\delta^{(0)}\neq0$ in Eq.~(\ref{eq:eff_pot_coefficient02_f})). In the bottom panels, we show 
the value of $r$ representing the phases,
the UN phase ($r=-1/2$), the D$_{2}$-BN phase ($-1<r<-1/2$), and the D$_{4}$-BN phase ($r=-1$). The cyan lines in the right panels indicate the first-order phase transition.}
\label{fig:phase_non_delta0_A0}
\end{center}
\end{figure}

\begin{figure}[tb]
\begin{center}
\includegraphics[scale=0.3]{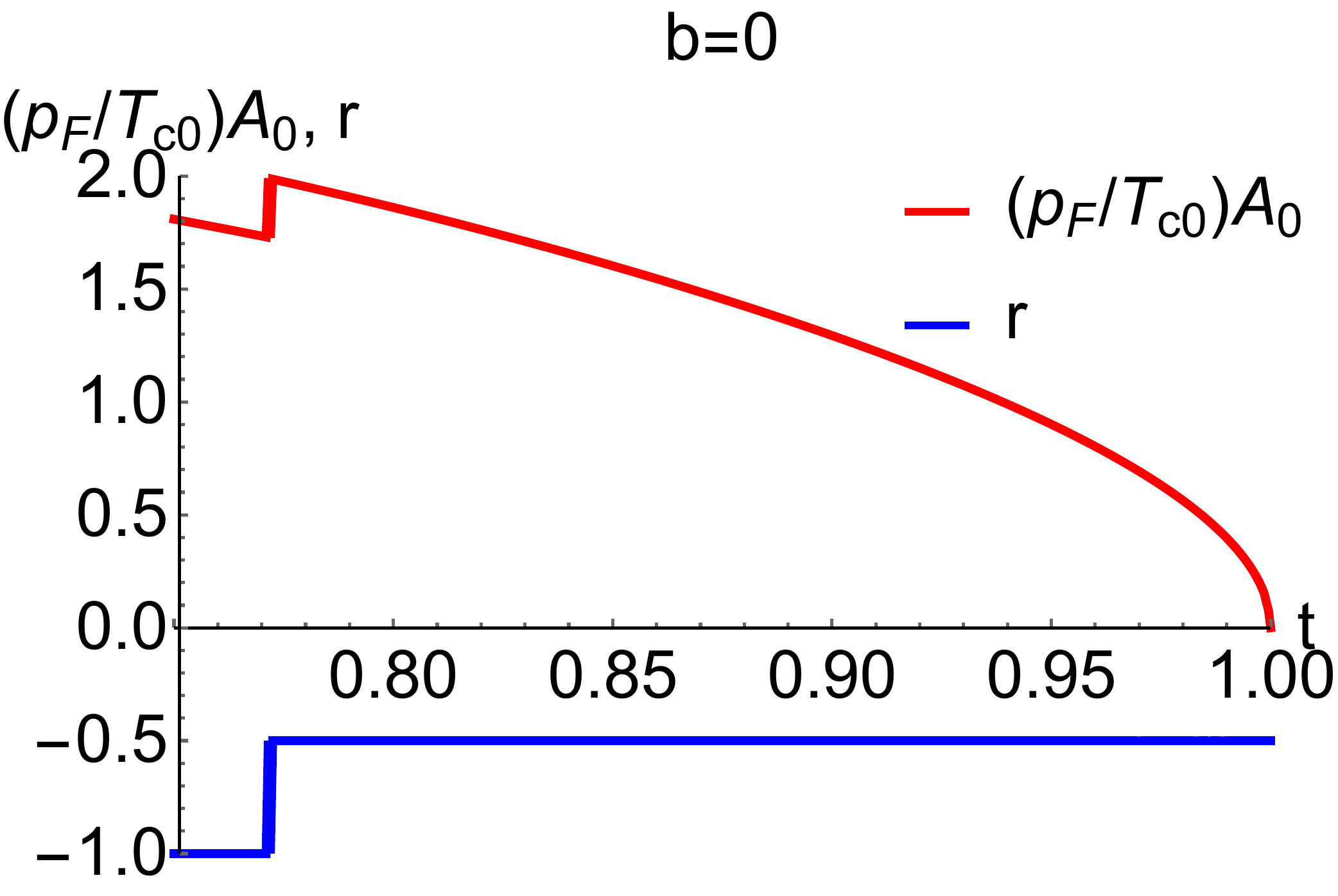}
\caption{The plots of $A_{0}$ and $r$ as functions of $t$ for $b=0$. The red line indicates $A_{0} \ge 0$, and the blue line indicates $-1 \le r \le -1/2$. For $b=0$, $A_{0}$ and $r$ are discontinuous at the first-order phase transition at $t=0.772$.}
\label{fig:Fig_A0_r_t_b0b}
\end{center}
\end{figure}

We plot the phase diagrams of $A_{0}$ and $r$ on the plane of the temperature ($t=T/T_{c0}$) and the magnetic field ($b=\gamma_{n}B/((1+F_{0}^{a})T_{c0})$) in Fig.~\ref{fig:phase_non_delta0_A0}.
Notice that $t$ and $b$ are dimensionless quantities.
In Figs.~\ref{fig:phase_non_delta0_A0}(a) and \ref{fig:phase_non_delta0_A0}(c), we show the previous results for the metastable states without the 8th order terms by setting $\delta_{0}=0$ in Eq.~\eqref{eq:eff_pot_coefficient02_f}.
In Figs.~\ref{fig:phase_non_delta0_A0}(b) and \ref{fig:phase_non_delta0_A0}(d), we show the new results with the 8th order terms by setting $\delta_{0}\neq0$ as in Eq.~\eqref{eq:eff_pot_coefficient0_parameters_FL_f}.
The upper and lower panels are for $A_{0}$ and $r$, respectively.
In the case of $\delta_{0}=0$, the UN phase appears at the temperature lower than the critical temperature ($t < 1$) and the zero magnetic field ($b=0$), and
the D$_{2}$-BN phase appears at the finite magnetic field, and it turns to be the D$_{4}$-BN phase at the stronger magnetic field. 
The $\delta_{0}\neq0$ case has almost no difference to the $\delta_{0}=0$ case around the critical temperature ($t\approx1$).
However, the particular difference appears at the lower temperature and the smaller magnetic field.
In the $\delta_{0}=0$ case, the UN phase and the D$_{2}$-BN phase are almost unchanged from the higher temperature to the lower temperature for the fixed magnetic field.
In contrast, in the $\delta_{0}\neq0$ case, the UN and D$_{2}$-BN phases disappear at the low temperature, and the D$_{4}$-BN phase appears.
This change is the first-order phase transition.
The transition lines are shown by the cyan lines in Fig.~\ref{fig:phase_non_delta0_A0}.
The critical temperature at zero magnetic field is $t_{c}=0.772$.
The existence of the first-order phase transition is also shown in Fig.~\ref{fig:Fig_A0_r_t_b0b}.
The discontinuous changes in $A_{0}$ and $r$ at $t_{c}=0.772$ for the zero magnetic field ($b=0$) can be seen clearly.
The first-order phase transition has not been known in the analysis of the GL theory 
up to the 6th order for the neutron $^{3}P_{2}$ superfluidity, while it was found in the analysis based on 
the BdG equation with quasi-classical approximation~\cite{Mizushima:2016fbn}.
Thus,  the 8th order terms for $\delta_{0}\neq 0$ in the GL equation 
capture all phase transitions in the neutron $^{3}P_{2}$ superfluidity at least qualitatively.

\section{Summary and discussion}

We have discussed the GL free energy density for the neutron $^{3}P_{2}$ superfluidity in terms of the $LS$ potential between two neutrons.
Adopting the path-integral formalism, the auxiliary field, and the Nambu-Gor'kov representation,
we have derived the GL free energy density up to the 8th order terms of the condensate.
We also have presented the detailed description of the calculation.
We have found that the 8th order terms support the ground state as the global minimum, and it is practically useful for the variational calculation for the GL free energy density. 
This resolves the problem of the GL expansions up to the 6th order in which 
the system is unstable for large values of the order parameters. 
We have shown the phase diagram of the nematic phases, {\it i.e.}, the UN phase, the D$_{2}$-BN phase, and the D$_{4}$-BN phase on the plane spanned by the temperature and the magnetic field.
It has tuned out that the D$_{4}$-BN phase appears at the low temperature and the smaller magnetic field, 
and that there is a first-order phase transition between the D$_{4}$-BN and D$_{2}$-BN phases.  
Those properties were not known in the previous studies in the GL free energy density, 
in which the maximum order in the condensate was limited to the 6th order.
Therefore, the 8th order terms capture a qualitatively correct phase diagram for the neutron $^{3}P_{2}$ superfluidity.

We comment that it was known already in the BdG equation that the first-order phase transition exists at the low temperature and the finite magnetic field~\cite{Mizushima:2016fbn}.
When we compare our result with the one from the BdG equation,
however, we notice that the values of the temperature and the magnetic field in which the first-order phase transition appears are quite different:
the transition line attaches to the temperature axis in our result, while it attaches to the axis of the magnetic field in the analysis by the BdG equation.
We will report a further detailed analysis soon.

The existence of the first-order transition can affect the properties of the neutron stars.
For example, it is one of the interesting questions to ask how the quantum vortices are affected by the first-order phase transition.
For future work, 
a connection of the neutron $^{3}P_{2}$ phase to the other phases such as hyperon matter, quark matter and so on should be studied. 
When a neutron star rotates, a connection between Abelian quantum vortices in the hadron matter and 
non-Abelian quantum vortices (color magnetic flux tubes)
\cite{Balachandran:2005ev,Nakano:2007dr,Eto:2009kg}
in the quark matter was discussed 
in the presence or absence of 
boojums which are endpoints of vortices~\cite{Cipriani:2012hr,Alford:2018mqj,Chatterjee:2018nxe,Chatterjee:2019tbz,Cherman:2018jir}.
The boojums, if they exist, may affect dynamical properties of the neutron stars.
See Ref.~\cite{Eto:2013hoa} for a review of non-Abelian quantum vortices in quark matter.
It is also important to study cyclic and ferromagnetic phases, 
which are possible for general GL theory with total angular momentum two 
\cite{Mermin:1974zz,2010arXiv1001.2072K}.
The former makes it possible to realize one-third quantized non-Abelian vortices~\cite{Semenoff:2006vv} 
which form a network in collision~\cite{Kobayashi:2008pk}, while  
the latter could be relevant to the properties of the magnetic field in magnetars.
The cyclic and ferromagnetic phases are topological matter (Weyl semimetals) which can have gapless fermions in the bulk~\cite{Mizushima:2016fbn,Mizushima:2017pma}.
It is also an interesting to apply our formulation 
to superfluidity to higher spin systems~\cite{PhysRevX.8.011029}.

It is also an interesting question to investigate the higher-partial wave component, {\it i.e.}, $^{3}F_{2}$, which couples to the $^{3}P_{2}$ component through the tensor interaction~\cite{Amundsen:1984qc,Takatsuka:1992ga,Zverev:2003ak}.
The mixing between the $^{3}P_{2}$ component and the $^{3}F_{2}$ component induces the enhancement of the amplitude of the gap value and enhances the stability of the neutron superfluidity, and it can provide richer phases than the pure $^{3}P_{2}$ component~\cite{Zverev:2003ak}.
Those new phases should be studied in detail in future studies.

\section*{Acknowledgment}
This work is supported by the Ministry of Education, Culture, Sports, Science (MEXT)-Supported Program for the Strategic Research Foundation at Private Universities ``Topological Science" (Grant No. S1511006). 
C.~C. acknowledges support as an International Research Fellow of the Japan Society for the Promotion of Science (JSPS) (Grant No. 16F16322). 
This work is also supported in part by 
JSPS Grant-in-Aid for Scientific Research (KAKENHI Grant No. 16H03984 (M.~N.), No.~18H01217 (M.~N.), No.~17K05435 (S.~Y.), and No.~19K14713 (C.~C.)), and also by MEXT KAKENHI Grant-in-Aid for Scientific Research on Innovative Areas ``Topological Materials Science'' (Grant No.~15H05855 (M.~N.)).

\appendix

\section{Trace calculation}
\label{sec:trace_calculation}

We present some examples of the trace calculation of Eq.~\eqref{eq:eff_pot_weakcoupling_b}.
We demonstrate the calculations of the log term in Sec.~\ref{sec:log_term} 
and the calculation of the power series in Sec.~\ref{sec:power_series} to derive the GL free energy density \eqref{eq:eff_pot_coefficient02_f}.

\subsection{Log term}
\label{sec:log_term}

We follow the several steps in the calculation.
First, we express the trace, ${\mathrm{Tr}}$, by the state $|\tau,\vec{x}\rangle$ in the real space (imaginary time and three-dimensional space). 
Second, we insert the complete set in the momentum space whose basis is denoted by $| i\omega_{n}, \vec{p} \rangle$ with the Matsubara frequency $\omega_{n}$ and the three-dimensional momentum $\vec{p}$.
Third, we utilize the plane-wave wave function $\displaystyle \langle i\omega_{n}, \vec{p} | \tau,\vec{x} \rangle = e^{-i\omega_{n}\tau+i\vec{p}\cdot\vec{x}}$.
With the above steps, we perform the calculation of the logarithmic term as
\begin{eqnarray}
&&
  - \mathrm{Tr} \ln \bigl( S_{m}(\tau,\vec{x})^{-1} \bigr)
\nonumber \\
&=&
  - \mathrm{tr} \,
    \int \drm\tau \, \drm\vec{x} \,
    \langle \tau,\vec{x} |
    \ln
    \left(
    \begin{array}{cc}
        \partial_{\tau} - \dfrac{\vec{\nabla}^{2}}{2m} - \mu - \vec{\mu}_{n} \!\cdot\! \vec{B} & 0 \\
        0 & \partial_{\tau} + \dfrac{\vec{\nabla}^{2}}{2m} + \mu + \vec{\mu}_{n}^{t} \!\cdot\! \vec{B}
    \end{array}
    \right)
    | \tau,\vec{x} \rangle
\nonumber \\
&=&
  - \mathrm{tr} \,
    \int \drm\tau \, \drm\vec{x} \,
    \langle \tau,\vec{x} |
    \ln
    \left(
    \begin{array}{cc}
        \partial_{\tau} - \dfrac{\vec{\nabla}^{2}}{2m} - \mu - \vec{\mu}_{n} \!\cdot\! \vec{B} & 0 \\
        0 & \partial_{\tau} + \dfrac{\vec{\nabla}^{2}}{2m} + \mu + \vec{\mu}_{n}^{t} \!\cdot\! \vec{B}
    \end{array}
    \right)
    T \sum_{n} \int \frac{\drm^{3}\vec{p}}{(2\pi)^{3}} | i\omega_{n}, \vec{p} \rangle
    \langle i\omega_{n}, \vec{p} | \tau,\vec{x} \rangle
\nonumber \\
&=&
  - \beta V
    \int \frac{\drm^{3}\vec{p}}{(2\pi)^{3}}
    \Bigl( 
           2 \, T \sum_{n} \ln \bigl( -i\omega_{n} + \xi_{\vec{p}}^{+} + i\eta \bigr) \bigl( -i\omega_{n} + \xi_{\vec{p}}^{-} + i\eta \bigr)
        + 2 \, T \sum_{n} \ln \bigl( -i\omega_{n} -  \xi_{\vec{p}}^{+} - i\eta \bigr) \bigl( -i\omega_{n} -  \xi_{\vec{p}}^{-} - i\eta \bigr)
    \Bigr),\nonumber\\
\end{eqnarray}
where we have  introduced the Matsubara frequency $\omega_{n}=(2n+1)\pi/\beta$ ($n \in \mathbb{Z}$) with the inverse temperature $\beta=1/T$, and the particle energy $ \xi_{\vec{p}}^{\pm} \equiv \xi_{\vec{p}} \pm |\vec{\mu}_{n}| |\vec{B}|$ with $ \xi_{\vec{p}} \equiv  {\vec{p}^{2}}/({2m}) - \mu$.
The summation of the Matsubara frequencies are performed as
\begin{eqnarray}
  T \sum_{n}
  \ln \bigl( -i\omega_{n} + \xi_{\vec{p}}^{\pm} + i\eta \bigr)
&=&
  T \sum_{n}
  \int_{0}^{\xi_{\vec{p}}} \drm\xi \frac{1}{-i\omega_{n} + \xi + i\eta}
\nonumber \\
&=&
   \frac{1}{2} \xi_{\vec{p}}^{\pm}
+ \frac{1}{\beta} \ln\bigl( 1+e^{-\beta \xi_{\vec{p}}^{\pm}} \bigr)
- \frac{1}{\beta} \ln2,
\end{eqnarray}
where we have used the trick to introduce the integral about $\xi$ and have used the relation
\begin{eqnarray}
 \sum_{n} f(i\omega_{n})
=
 \frac{\beta}{2\pi i} \oint_{C} \drm p_{0} \, f(p_{0}) \frac{1}{2} \mathrm{tanh} \biggl( \frac{\beta}{2}p_{0} \biggr),
\label{eq:Matsubara_sum}
\end{eqnarray}
for an analytic function $f(p_{0})$.
The path $C$ turns around $i\omega_{n}$ counterclockwise.
In the end of the above calculation, we obtain
\begin{eqnarray}
  - \mathrm{Tr} \ln \bigl( S_{m}(\tau,\vec{x})^{-1} \bigr)
=
  - \beta V \,
    \frac{2}{\beta} \int \frac{\drm^{3}\vec{p}}{(2\pi)^{3}}
    \ln\bigl( 1+e^{-\beta \xi_{\vec{p}}^{\pm}} \bigr),
\end{eqnarray}
where the constant terms are neglected.
We also have neglected the term $\frac{1}{\beta} \ln\bigl( 1+e^{\beta \xi_{\vec{p}}^{\pm}} \bigr)$ which can be regarded as a shift by treating the particle and hole on the equal footing in the Nambu-Gor'kov representation.
By dividing the above quantity by the total volume $\displaystyle \int \drm\tau \, \drm^{3}\vec{x} \, 1 = \beta V$ ($V$ being the volume of the three-dimensional space) and 
by multiplying the factor $1/2$ for removing the double-counting in the Nambu-Gor'kov representation,
we obtain $f_{0}$ which appears in Eq.~\eqref{eq:eff_pot_free_magneticfield}.

\subsection{Power series}
\label{sec:power_series}

We apply the essentially same procedure of the calculation of the power series ($N$th order) in Eq.~\eqref{eq:eff_pot_weakcoupling_b}.
The different point is the necessity of the multiple number of states for the completeness relations, such as $|\tau_{1},\vec{x}_{1} \rangle$, $|\tau_{2},\vec{x}_{2} \rangle$, $\dots$, $|\tau_{N},\vec{x}_{N} \rangle$ in the real space and $|i\omega_{1},\vec{p}_{1}\rangle$, $|i\omega_{2},\vec{p}_{2}\rangle$, $\dots$, $|i\omega_{N},\vec{p}_{N}\rangle$ in the momentum space.
The $N$th order term is calculated as follows:
\begin{eqnarray}
&&
  \mathrm{Tr} \Bigl( \bigl( S_{m}(\tau,\vec{x}) \hat{A}(\tau,\vec{x}) \bigr)^{N} \Bigr)
\nonumber \\
&=&
  \mathrm{tr}
  \int \drm\tau \, \drm^{3}\vec{x} \,
  \langle \tau, \vec{x} | \bigl( S_{m}(\tau,\vec{x}) \hat{A}(\tau,\vec{x}) \bigr)^{N} | \tau, \vec{x} \rangle
\nonumber \\
&=&
  \mathrm{tr}
  \int \drm\tau \, \drm^{3}\vec{x} \,
  \langle \tau, \vec{x} |
  S_{m}(\tau,\vec{x}) \hat{A}(\tau,\vec{x}) \cdots
  S_{m}(\tau,\vec{x}) \hat{A}(\tau,\vec{x})
  S_{m}(\tau,\vec{x}) \hat{A}(\tau,\vec{x})
  | \tau, \vec{x} \rangle
\nonumber \\ \nonumber \\
&=&
  \mathrm{tr}
  \int \drm\tau_{1} \, \drm^{3}\vec{x}_{1} \,
  \langle \tau_{1}, \vec{x}_{1} |
  S_{m}(\tau_{N},\vec{x}_{N}) \hat{A}(\tau_{N},\vec{x}_{N})
  \int \drm\tau_{N} \, \drm^{3}\vec{x}_{N} \,
  | \tau_{N}, \vec{x}_{N} \rangle \langle \tau_{N}, \vec{x}_{N} |
  \cdots
  \nonumber \\ && \times 
  S_{m}(\tau_{2},\vec{x}_{2}) \hat{A}(\tau_{2},\vec{x}_{2})
  \int \drm\tau_{2} \, \drm^{3}\vec{x}_{2} \,
  | \tau_{2}, \vec{x}_{2} \rangle \langle \tau_{2}, \vec{x}_{2} |
  S_{m}(\tau_{1},\vec{x}_{1}) \hat{A}(\tau_{1},\vec{x}_{1}) | \tau_{1}, \vec{x}_{1} \rangle
\nonumber \\ \nonumber \\
&=&
  \mathrm{tr}
  \int \drm\tau_{1} \, \drm^{3}\vec{x}_{1}
  \int \drm\tau_{2} \, \drm^{3}\vec{x}_{2} \,
  \cdots
  \int \drm\tau_{N} \, \drm^{3}\vec{x}_{N} \,
  \nonumber \\ && \times 
  \langle \tau_{1}, \vec{x}_{1} |
  S_{m}(\tau_{N},\vec{x}_{N}) \hat{A}(\tau_{N},\vec{x}_{N})
  | \tau_{N}, \vec{x}_{N} \rangle \langle \tau_{N}, \vec{x}_{N} |
  \cdots
  S_{m}(\tau_{2},\vec{x}_{2}) \hat{A}(\tau_{2},\vec{x}_{2})
  | \tau_{2}, \vec{x}_{2} \rangle \langle \tau_{2}, \vec{x}_{2} |
  S_{m}(\tau_{1},\vec{x}_{1}) \hat{A}(\tau_{1},\vec{x}_{1}) | \tau_{1}, \vec{x}_{1} \rangle
\nonumber \\ \nonumber \\
&=&
  \mathrm{tr}
  \int \drm\tau_{1} \, \drm^{3}\vec{x}_{1}
  \int \drm\tau_{2} \, \drm^{3}\vec{x}_{2} \,
  \cdots
  \int \drm\tau_{N} \, \drm^{3}\vec{x}_{N} \,
  \nonumber \\ && 
  \langle \tau_{1}, \vec{x}_{1} |
  S_{m}(\tau_{N},\vec{x}_{N}) \hat{A}(\tau_{N},\vec{x}_{N})
  T \sum_{n_{N}} \int \frac{\drm^{3}\vec{p}_{N}}{(2\pi)^{3}}
  | i\omega_{n_{N}}, \vec{p}_{N} \rangle
  \langle i\omega_{n_{N}}, \vec{p}_{N} | \tau_{N}, \vec{x}_{N} \rangle
  \nonumber \\ && \times
  \cdots
  \nonumber \\ && \times
  \langle \tau_{3}, \vec{x}_{3} |
  S_{m}(\tau_{2},\vec{x}_{2}) \hat{A}(\tau_{2},\vec{x}_{2})
  T \sum_{n_{2}} \int \frac{\drm^{3}\vec{p}_{2}}{(2\pi)^{3}}
  | i\omega_{n_{2}}, \vec{p}_{2} \rangle
  \langle i\omega_{n_{2}}, \vec{p}_{2} | \tau_{2}, \vec{x}_{2} \rangle
  \nonumber \\ && \times
  \langle \tau_{2}, \vec{x}_{2} |
  S_{m}(\tau_{1},\vec{x})_{1} \hat{A}(\tau_{1},\vec{x}_{1})
  T \sum_{n_{1}} \int \frac{\drm^{3}\vec{p}_{1}}{(2\pi)^{3}}
  | i\omega_{n_{1}}, \vec{p}_{1} \rangle
  \langle i\omega_{n_{1}}, \vec{p}_{1} | \tau_{1}, \vec{x}_{1} \rangle
\nonumber \\ \nonumber \\
&=&
  \mathrm{tr}
  \int \drm\tau_{1} \, \drm^{3}\vec{x}_{1}
  \int \drm\tau_{2} \, \drm^{3}\vec{x}_{2}
  \cdots
  \int \drm\tau_{N} \, \drm^{3}\vec{x}_{N}
  \nonumber \\ && 
  \langle \tau_{1}, \vec{x}_{1} | i\omega_{n_{N}}, \vec{p}_{N} \rangle
  S_{m}(\tau_{N},\vec{x}_{N}) \hat{A}(\tau_{N},\vec{x}_{N})
  T \sum_{n_{N}} \int \frac{\drm^{3}\vec{p}_{N}}{(2\pi)^{3}}  
  e^{-i\omega_{n_{N}}\tau_{N}+i\vec{p}_{N}\cdot\vec{x}_{N}}
  \nonumber \\ && \times 
  \cdots
  \nonumber \\ && \times 
  \langle \tau_{3}, \vec{x}_{3} | i\omega_{n_{2}}, \vec{p}_{2} \rangle
  S_{m}(\tau_{2},\vec{x}_{2}) \hat{A}(\tau_{2},\vec{x}_{2})
  T \sum_{n_{2}} \int \frac{\drm^{3}\vec{p}_{2}}{(2\pi)^{3}}  
  e^{-i\omega_{n_{2}}\tau_{2}+i\vec{p}_{2}\cdot\vec{x}_{2}}
  \nonumber \\ && \times 
  \langle \tau_{2}, \vec{x}_{2} | i\omega_{n_{1}}, \vec{p}_{1} \rangle
  S_{m}(\tau_{1},\vec{x}_{1}) \hat{A}(\tau_{1},\vec{x}_{1})
  T \sum_{n_{1}} \int \frac{\drm^{3}\vec{p}_{1}}{(2\pi)^{3}}
  e^{-i\omega_{n_{1}}\tau_{1}+i\vec{p}_{1}\cdot\vec{x}_{1}}
\nonumber \\ \nonumber \\
&=&
  \mathrm{tr}
  \int \drm\tau_{1} \, \drm^{3}\vec{x}_{1}
  \int \drm\tau_{2} \, \drm^{3}\vec{x}_{2}
  \cdots
  \int \drm\tau_{N} \, \drm^{3}\vec{x}_{N} \,
  \nonumber \\ && 
  e^{i\omega_{n_{N}}\tau_{1}-i\vec{p}_{N}\cdot\vec{x}_{1}}
  \bar{S}_{m}(i\omega_{n_{N}},\vec{p}_{N}) \hat{\bar{A}}(\tau_{N},\vec{x}_{N};\vec{p}_{N})
  T \sum_{n_{N}} \int \frac{\drm^{3}\vec{p}_{N}}{(2\pi)^{3}}  
  e^{-i\omega_{n_{N}}\tau_{N}+i\vec{p}_{N}\cdot\vec{x}_{N}}
  \nonumber \\ && \times 
  \cdots
  \nonumber \\ && \times 
  e^{i\omega_{n_{2}}\tau_{3}-i\vec{p}_{2}\cdot\vec{x}_{3}}
  \bar{S}_{m}(i\omega_{n_{2}},\vec{p}_{2}) \hat{\bar{A}}(\tau_{2},\vec{x}_{2};\vec{p}_{2})
  T \sum_{n_{2}} \int \frac{\drm^{3}\vec{p}_{2}}{(2\pi)^{3}}  
  e^{-i\omega_{n_{2}}\tau_{2}+i\vec{p}_{2}\cdot\vec{x}_{2}}
  \nonumber \\ && \times 
  e^{i\omega_{n_{1}}\tau_{2}-i\vec{p}_{1}\cdot\vec{x}_{2}}
  \bar{S}_{m}(i\omega_{n_{1}},\vec{p}_{1}) \hat{\bar{A}}(\tau_{1},\vec{x}_{1};\vec{p}_{1})
  T \sum_{n_{1}} \int \frac{\drm^{3}\vec{p}_{1}}{(2\pi)^{3}}
  e^{-i\omega_{n_{1}}\tau_{1}+i\vec{p}_{1}\cdot\vec{x}_{1}}
\nonumber \\ \nonumber \\
&=&
  T \sum_{n_{1}} \int \frac{\drm^{3}\vec{p}_{1}}{(2\pi)^{3}} \,
  T \sum_{n_{2}} \int \frac{\drm^{3}\vec{p}_{2}}{(2\pi)^{3}} \,
  \cdots
  T \sum_{n_{N}} \int \frac{\drm^{3}\vec{p}_{N}}{(2\pi)^{3}} \,
  \int \drm\tau_{1} \, \drm^{3}\vec{x}_{1}
  \int \drm\tau_{2} \, \drm^{3}\vec{x}_{2}
  \cdots
  \int \drm\tau_{N} \, \drm^{3}\vec{x}_{N}
  \nonumber \\ && 
  e^{-i(\omega_{n_{N}}-\omega_{n_{N-1}})\tau_{N}}
  \cdots
  e^{-i(\omega_{n_{2}}-i\omega_{n_{1}})\tau_{2}}
  e^{-i(\omega_{n_{1}}-\omega_{n_{N}})\tau_{1}}
  \times
  e^{i(\vec{p}_{N}-\vec{p}_{N-1})\cdot\vec{x}_{N}}
  \cdots
  e^{i(\vec{p}_{2}-\vec{p}_{1})\cdot\vec{x}_{2}}
  e^{i(\vec{p}_{1}-\vec{p}_{N})\cdot\vec{x}_{1}}
  \nonumber \\ && \times 
  \mathrm{tr}
  \Bigl(
        \bar{S}_{m}(i\omega_{n_{N}},\vec{p}_{N}) \hat{\bar{A}}(\tau_{n_{N}},\vec{x}_{N};\vec{p}_{N})
        \cdots
        \bar{S}_{m}(i\omega_{n_{2}},\vec{p}_{1}) \hat{\bar{A}}(\tau_{2},\vec{x}_{2};\vec{p}_{2})
        \bar{S}_{m}(i\omega_{n_{1}},\vec{p}_{1}) \hat{\bar{A}}(\tau_{1},\vec{x}_{1};\vec{p}_{1})
  \Bigr),
\end{eqnarray}
where we have used the completeness relations
\begin{eqnarray}
   \int \drm\tau_{i} \, \drm^{3}\vec{x}_{i} \, | \tau_{i}, \vec{x}_{i} \rangle \langle \tau_{i}, \vec{x}_{i} |=1,
\end{eqnarray}
in the real space and
\begin{eqnarray}
   T \sum_{n_{i}} \int \frac{\drm^{3}\vec{p}_{i}}{(2\pi)^{3}} | i\omega_{n_{i}}, \vec{p}_{i} \rangle \langle i\omega_{n_{i}}, \vec{p}_{i} |=1,
\end{eqnarray}
in the momentum space with the Matsubara frequency $\omega_{n_{i}}=(2n_{i}+1)\pi/\beta$ for each $i=1,2,\dots,N$.
The trace, $\mathrm{tr}$, is taken over for the spin of the neutron and the up and down-components in the Nambu-Gor'kov representation.
In the above equations, we have introduced the propagator of a nucleon under the finite magnetic field
\begin{eqnarray}
  \bar{S}_{m}(i\omega_{n},\vec{p})^{-1}
\equiv
   \left(
    \begin{array}{cc}
          -i\omega_{n} + \xi_{\vec{p}} - \vec{\mu}_{n}\!\cdot\!\vec{B} + i\eta & 0 \\
          0 & -i\omega_{n} - \xi_{\vec{p}} + \vec{\mu}_{n}^{t}\!\cdot\!\vec{B} - i\eta
    \end{array}
    \right),
\end{eqnarray}
and the matrix of the auxiliary field
\begin{eqnarray}
  \hat{\bar{A}}(\tau,\vec{x}; \vec{p})
&\equiv&
   \left(
    \begin{array}{cc}
          0 & {\displaystyle \sum_{a,b}} A^{ab}(\tau,\vec{x}) \bar{t}^{ab\dag}(\vec{p}) \\
          {\displaystyle \sum_{a,b}} A^{ab\dag}(\tau,\vec{x}) \bar{t}^{ab}(\vec{p}) & 0
    \end{array}
    \right),
\label{eq:taubarhat_def}
\\
 \bar{t}^{ab}(\vec{p})
&\equiv&
 -i \biggl( \frac{1}{2}\Sigma^{a\dag} p^{b} + \frac{1}{2}\Sigma^{b\dag} p^{a} - \frac{1}{3}\delta^{ab}\sum_{c}\Sigma^{c\dag} p^{c} \biggr),
\label{eq:tbar_def}
\end{eqnarray}
which are (partially) represented in the momentum space and the Matsubara frequency.
The imaginary time and the positions $(\tau_{i},\vec{x}_{i})$ are different for $i=1,2,\dots,N$.
As for the kinetic term with $N=2$, we use the expansion 
\begin{eqnarray}
   \hat{\bar{A}}(\tau_{2},\vec{x}_{2};\vec{p}_{2})
&=&
   \hat{\bar{A}}(\tau_{1},\vec{x}_{1};\vec{p}_{2})  
+ (\tau_{2}-\tau_{1})\partial_{\tau}\hat{\bar{A}}(\tau,\vec{x}_{1};\vec{p}_{2}) \Bigr|_{\tau=\tau_{1}}
+ \frac{1}{2} (\tau_{2}-\tau_{1})^{2} \partial_{\tau}^{2} \hat{\bar{A}}(\tau,\vec{x}_{1};\vec{p}_{2}) \Bigr|_{\tau=\tau_{1}}
   \nonumber \\ &&
+ (\vec{x}_{2}-\vec{x}_{1}) \!\cdot\! \vec{\nabla}_{x} \hat{\bar{A}}(\tau_{1},\vec{x};\vec{p}_{2}) \Bigr|_{\vec{x}=\vec{x}_{1}}
+ \frac{1}{2} \sum_{i,j} (x_{2}-x_{1})_{i} (x_{2}-x_{1})_{j} \nabla_{xi} \nabla_{xj} \hat{\bar{A}}(\tau_{1},\vec{x};\vec{p}_{2}) \Bigr|_{\vec{x}=\vec{x}_{1}}
+ \cdots,
\end{eqnarray}
by assuming that the difference between $(\tau_{2},\vec{x}_{2})$ and $(\tau_{1},\vec{x}_{1})$ is small.
The higher order terms in the derivatives are neglected.
In the case of $N\ge3$, we regard that all the different points $(\tau_{i},\vec{x}_{i})$ are identical to $(\tau,\vec{x})$, and apply the approximation
\begin{eqnarray}
   \hat{\bar{A}}(\tau_{i},\vec{x}_{i}; \vec{p}_{i}) \approx \hat{\bar{A}}(\tau,\vec{x}; \vec{p}_{i}),
\end{eqnarray}
for each $i=1,2,\dots,N$.
We also adopt the expansion for the magnetic field $\vec{B}$ up to the appropriate order, such as
\begin{eqnarray}
&&
 \frac{1}{ - i\omega_{n} + \xi_{\vec{p}} - \vec{\mu}_{n}\!\cdot\!\vec{B} }
\nonumber \\ 
&=&
       \frac{1}{ - i\omega_{n} + \xi_{\vec{p}} }
    + \frac{1}{\bigl( - i\omega_{n} + \xi_{\vec{p}}\bigr)^{2}}
       \vec{\mu}_{n}\!\cdot\!\vec{B}
    + \frac{1}{\bigl( - i\omega_{n} + \xi_{\vec{p}}\bigr)^{3}}
       \bigl(\vec{\mu}_{n}\!\cdot\!\vec{B}\bigr)^{2}
    + \frac{1}{\bigl( - i\omega_{n} + \xi_{\vec{p}}\bigr)^{4}}
       \bigl(\vec{\mu}_{n}\!\cdot\!\vec{B}\bigr)^{3}
    + \frac{1}{\bigl( - i\omega_{n} + \xi_{\vec{p}}\bigr)^{5}}
       \bigl(\vec{\mu}_{n}\!\cdot\!\vec{B}\bigr)^{4}
       \nonumber \\ && \hspace{0em}
    + {\cal O}\Bigl(\bigl(\vec{\mu}_{n}\!\cdot\!\vec{B}\bigr)^{5}\Bigr),
\end{eqnarray}
whenever it is necessary.

We leave comments on the tips for the integrals for the three-dimensional momentum.
Parametrizing $\vec{p}=(p\sin\theta\cos\varphi,p\sin\theta\sin\varphi,p\cos\theta)$ in the polar coordinate and assuming the isotropy in the momentum space, we calculate the polar integrals as
\begin{eqnarray}
   \int \drm\Omega_{\vec{p}} \, p^{a_{1}}p^{a_{2}} f(p)
&=& \frac{4\pi}{3} p^{2} \delta^{a_{1}a_{2}} f(p),
\nonumber \\ 
   \int \drm\Omega_{\vec{p}} \, p^{a_{1}}p^{a_{2}}p^{a_{3}}p^{a_{4}} f(p)
&=& \frac{4\pi}{15} p^{4} \bigl( \delta^{a_{1}a_{2}}\delta^{a_{3}a_{4}} + \delta^{a_{1}a_{3}}\delta^{a_{2}a_{4}} + \delta^{a_{1}a_{4}}\delta^{a_{2}a_{3}} \bigr) f(p),
\nonumber \\ 
 \int \drm\Omega_{\vec{p}} \, p^{a_{1}}p^{a_{2}}p^{a_{3}}p^{a_{4}}p^{a_{5}}p^{a_{6}} f(p)
&=&
  \frac{4\pi}{105} p^{6}
  \delta^{(a_{1}a_{2}} \delta^{a_{3}a_{4}} \delta^{a_{5}a_{6})} f(p),
\nonumber \\ 
   \int \drm\Omega_{\vec{p}} \, p^{a_{1}} p^{a_{2}} p^{a_{3}} p^{a_{4}} p^{a_{5}} p^{a_{6}} p^{a_{7}} p^{a_{8}} f(p)
&=&
   \frac{4\pi}{945} p^{8}
   \delta^{(a_{1}a_{2}} \delta^{a_{3}a_{4}} \delta^{a_{5}a_{6}} \delta^{a_{7}a_{8})} f(p),
\label{eq:replacement_p_isotropy}
\end{eqnarray}
with $\drm \Omega_{\vec{p}}=\sin\theta\drm\theta\drm\varphi$, where $f(p)$ is a function of $p=|\vec{p}|$.
We use the notation that $S^{a_{1}a_{2}\dots a_{n}}$ indicates the totally symmetric tensor for $a_{1},a_{2},\dots,a_{n}$ ($a_{i}=1,2,3$ for $i=1,2,\dots,n$), such as
\begin{eqnarray}
\delta^{(a_{1}a_{2}} \delta^{a_{3}a_{4}} \delta^{a_{5}a_{6})}
&=&
         \delta^{a_{1}a_{2}} \delta^{a_{3}a_{4}} \delta^{a_{5}a_{6}}
      + \delta^{a_{1}a_{2}} \delta^{a_{3}a_{5}} \delta^{a_{4}a_{6}}
      + \delta^{a_{1}a_{2}} \delta^{a_{3}a_{6}} \delta^{a_{4}a_{5}}
      + \delta^{a_{1}a_{3}} \delta^{a_{2}a_{4}} \delta^{a_{5}a_{6}}
         \nonumber \\ && \hspace{0em} 
      + \delta^{a_{1}a_{3}} \delta^{a_{2}a_{5}} \delta^{a_{4}a_{6}}
      + \delta^{a_{1}a_{3}} \delta^{a_{2}a_{6}} \delta^{a_{4}a_{5}}
      + \delta^{a_{1}a_{4}} \delta^{a_{2}a_{3}} \delta^{a_{5}a_{6}}
      + \delta^{a_{1}a_{4}} \delta^{a_{2}a_{5}} \delta^{a_{3}a_{6}}
         \nonumber \\ && \hspace{0em} 
      + \delta^{a_{1}a_{4}} \delta^{a_{2}a_{6}} \delta^{a_{3}a_{5}}
      + \delta^{a_{1}a_{5}} \delta^{a_{2}a_{3}} \delta^{a_{4}a_{6}}
      + \delta^{a_{1}a_{5}} \delta^{a_{2}a_{4}} \delta^{a_{3}a_{6}}
      + \delta^{a_{1}a_{5}} \delta^{a_{2}a_{6}} \delta^{a_{3}a_{4}}
         \nonumber \\ && \hspace{0em} 
      + \delta^{a_{1}a_{6}} \delta^{a_{2}a_{3}} \delta^{a_{4}a_{5}}
      + \delta^{a_{1}a_{6}} \delta^{a_{2}a_{4}} \delta^{a_{3}a_{5}}
      + \delta^{a_{1}a_{6}} \delta^{a_{2}a_{5}} \delta^{a_{3}a_{4}}.
\end{eqnarray}
As for the integrals for $p$, we adopt the quasiclassical approximation
\begin{eqnarray}
   \frac{1}{2\pi^{2}} \int f(p) p^{2}\,\drm p
\simeq
   N(0)
   \int f(p) \drm p
\simeq
   N(0)
   \int_{-\infty}^{\infty} f\bigl(2m\sqrt{\xi_{\vec{p}}+\mu}\bigr) \drm\xi_{\vec{p}},
\label{eq:Fermi_surface_approximation}
\end{eqnarray}
with $N(0)={m\,p_{F}}/({2\pi^{2}})$ being the number of the state density at the Fermi surface,
where we measure the energy from the chemical potential as $\xi_{\vec{p}}=\vec{p}^{2}/(2m)-\mu$.
In the first approximate equality, we have replaced the measure part of the momentum integral from $p^{2}/(2\pi^{2})$ to $p_{F}^{2}/(2\pi^{2})$ with the Fermi momentum $p_{F}$, and in the second approximate equality, we have introduced $\xi_{\vec{p}}$ instead of $p$ and have determined the integral range by $-\infty < \xi_{\vec{p}} < \infty$.
Such an extension of the integral range will be reasonable at low temperature.
Now we take the summation over the Matsubara frequencies $\omega_{n}$ and perform the momentum integrals for $p$.
For the term linear to $\mathrm{tr}\,A^{\ast}A$, we consider
\begin{eqnarray}
  T \sum_{n}
  \frac{1}{(i\omega_{n})^{2} - \xi_{\vec{p}}^{2}}
=
 -
 \frac{1}{2\xi_{\vec{p}}}
 \mathrm{tanh} \biggl( \frac{\beta}{2} \xi_{\vec{p}} \biggr),
 \label{eq:Matsubara_sum_2}
\end{eqnarray}
and obtain
\begin{eqnarray}
   \int \frac{\drm^{3}\vec{p}}{(2\pi)^{3}} \,
   \frac{\vec{p}^{2}}{3} \frac{1}{2\xi_{\vec{p}}}
   \mathrm{tanh} \biggl( \frac{\beta}{2} \xi_{\vec{p}} \biggr)
&\simeq&
   N(0)
   \frac{p_{F}^{2}}{3} \frac{1}{2}
   \int_{-D/2T}^{D/2T} \drm x \,
   \frac{\mathrm{tanh}\,x}{x}
\nonumber \\
&=&
   N(0)
   \frac{p_{F}^{2}}{3} \frac{1}{2}
   \Biggl(
         \bigl(\mathrm{tanh}\,x\bigr) \log|x| \Bigr|_{x=-D/2T}^{x=D/2T}
       - \int_{-D/2T}^{D/2T} \drm x \,
         \biggl(\frac{d}{\drm x}\mathrm{tanh}\,x\biggr) \log|x|
   \Biggr)
\nonumber \\
&\simeq&
   N(0)
   \frac{p_{F}^{2}}{3} \frac{1}{2}
   \Biggl(
         2
         \biggl(\mathrm{tanh}\,\frac{D}{2T}\biggr) \log\frac{D}{2T}
       - \int_{-\infty}^{\infty} \drm x \,
         \bigl(\mathrm{sech}\,x\bigr)^{2} \log|x|
   \Biggr)
\nonumber \\
&=&
   -N(0)
   \frac{p_{F}^{2}}{3}
   \log \frac{4e^{\gamma}}{\pi} \frac{2T}{D}.
\end{eqnarray}
Here, we have introduced the momentum cutoff $D$ above and below the Fermi surface and have used the integral
\begin{eqnarray}
   \int_{-\infty}^{\infty} \drm x \,
   \bigl(\mathrm{sech}\,x\bigr)^{2} \log|x|
= 2 \log \frac{4e^{\gamma}}{\pi},
\end{eqnarray}
with the Euler's constant $\gamma$.
For the other terms, we adopt the following procedure
\begin{eqnarray}
   \int_{-\infty}^{\infty} \drm \xi_{\vec{p}} \sum_{n=-\infty}^{\infty} \frac{1}{\bigl(\xi_{\vec{p}}^{2}+\omega_{n}^{2}\bigr)^{k}}
&=&
   \frac{\sqrt{\pi}\,\Gamma(k-1/2)}{\Gamma(k)} \sum_{n=-\infty}^{\infty} \frac{1}{\omega_{n}^{2k-1}}
\nonumber \\ 
&=&
   \frac{\sqrt{\pi}\,\Gamma(k-1/2)}{\Gamma(k)}
   \frac{2}{(\pi \, T)^{2k-1}}
   \biggl( 1-\frac{1}{2^{2k-1}} \biggr)
   \sum_{n=1}^{\infty} \frac{1}{n^{2k-1}}
\nonumber \\ 
&=&
   \frac{\sqrt{\pi}\,\Gamma(k-1/2)}{\Gamma(k)}
   \frac{2}{(\pi \, T)^{2k-1}}
   \biggl( 1-\frac{1}{2^{2k-1}} \biggr)
   \zeta(2k-1),
\end{eqnarray}
for a positive integer $k$.
Notice that the momentum integral is convergent without introducing the finite momentum cutoff $D$.
We have used $\omega_{n}=(2n+1)\pi \, T$ for the temperature $T$, the Gamma function $\Gamma(x)$, and the zeta function $\zeta(x)=\sum_{n=1}^{\infty}1/n^{x}$.

Finally, we comment that the equations represented by trace, $\mathrm{tr}$, for the symmetric, traceless, and complex matrix $A$ with $3\times3$ dimension have some identity relations.
For $N=4$, we have the identity~\cite{Sauls:1978lna}
\begin{eqnarray}
    \frac{1}{2} \mathrm{tr}(AA) \, \mathrm{tr}(A^{\ast}A^{\ast})
 + \mathrm{tr}(AA^{\ast}) \, \mathrm{tr}(AA^{\ast})
  - \mathrm{tr}(AA^{\ast}AA^{\ast})
 - 2 \mathrm{tr}(AAA^{\ast}A^{\ast})
= 0,
\end{eqnarray}
and for $N=6$ we have the identities
\begin{eqnarray}
 \mathrm{tr} \bigl( A A A^{\ast} A A^{\ast} A^{\ast} \bigr)
&=&
 \mathrm{tr}\bigl( A A A^{\ast} A^{\ast} A A^{\ast} \bigr),
\nonumber \\ 
 \mathrm{tr}\bigl( A A \bigr)
 \mathrm{tr}\bigl( A A^{\ast} A^{\ast} A^{\ast} \bigr)
&=&
 \mathrm{tr}\bigl( A^{\ast} A^{\ast} \bigr)
 \mathrm{tr}\bigl( A A A A^{\ast} \bigr),
\end{eqnarray}
which are sometimes useful to reduce the number of the terms in the final result.
In the end of the above calculations, we obtain the GL free energy density \eqref{eq:eff_pot_B2w4_coefficient02_f}.

\section{Symmetries}
\label{sec:symmetries}
In this Appendix, 
we summarize the symmetries of the order parameter $A$ parametrized in Eq.~\eqref{eq:A_gs}.
We remember that the symmetry of $A$ is given
\begin{eqnarray}
 A(\tau,\vec{x}) \rightarrow e^{i\alpha} O(\theta,\vec{n}) A(\tau,\tilde{\vec{x}}) O^{t}(\theta',\vec{n}'),
\end{eqnarray}
in the Lagrangian,
where $e^{i\alpha} \in \mathrm{U}(1)$ and $O(\theta,\vec{n})$, $O(\theta',\vec{n}') \in \mathrm{SO}(3)$ with $\vec{n}$ ($\vec{n}'$) the rotation axis and $\theta$ ($\theta'$) the rotation angle around $\vec{n}$ ($\vec{n}'$).
We notice that $O(\theta,\vec{n})$ is the rotation in the spin space, and $O(\theta',\vec{n}')$ is the rotation in the real space.
$\tilde{\vec{x}}$ is the vector rotated by $O(\theta',\vec{n}')$ from $\vec{x}$.
The above symmetries are spontaneously broken to subgroups in general in the nematic phase in Eq.~\eqref{eq:A_gs}.
Instead, there are the U(1) symmetry, the D$_{2}$ symmetry, and the D$_{4}$ symmetry, which are the symmetries in the UN phase ($r=-1/2$), the D$_{2}$-BN phase ($-1<r<-1/2$), and the D$_{4}$-BN phase ($r=-1$).
We will explain the concrete forms of those symmetries in the following.

\subsection{UN phase ($r=-1/2$)}

For $r=-1/2$, the order parameter $A(\tau,\vec{x})$ is expressed as
\begin{eqnarray}
  A_{\mathrm{UN}}(\tau,\vec{x})
=
A_{0}
\left(
\begin{array}{ccc}
 -1/2 & 0  & 0  \\
 0 & -1/2  & 0  \\
 0 & 0 & 1  
\end{array}
\right).
\label{eq:tau_gs_UN}
\end{eqnarray}
This is invariant under the rotation around the $z$-axis
\begin{eqnarray}
  A_{\mathrm{UN}}(\tau,\vec{x})
\rightarrow
  O(\theta) A_{\mathrm{UN}}(\tau,\tilde{\vec{x}}) O^{t}(\theta),
\end{eqnarray}
with $e^{i\alpha}=1$ ($\alpha=0$),
where $O(\theta) \in \mathrm{O}(2)$ is the rotation operator
\begin{eqnarray}
O(\theta)
=
\left(
\begin{array}{ccc}
 \cos \theta & -\sin \theta  & 0  \\
 \sin \theta & \cos \theta  & 0  \\
 0 & 0  & 1   
\end{array}
\right),
\end{eqnarray}
with the rotation angle $\theta$ ($0\le\theta<2\pi$).
This phase is called the uniaxial nematic (UN) phase, because the remaining symmetry is $\mathrm{O}(2) \approx \mathrm{U}(1)$.
We notice that the rotation in the spin space and the rotation in the real space is locked to each other, $O(\theta,\vec{n})=O(\theta',\vec{n}')$ with $\vec{n}=\vec{n}'$ and $\theta=\theta'$. The locking also occurs for the BN phase as we will explain in the next subsection.

\subsection{D$_{2}$-BN phase ($-1< r < -1/2$)}

For $-1< r < -1/2$, the order parameter $A(\tau,\vec{x})$ is expressed as
\begin{eqnarray}
  A_{\mathrm{D}_{2}\mathrm{BN}}(\tau,\vec{x})
=
A_{0}
\left(
\begin{array}{ccc}
 -r & 0  & 0  \\
 0 & -1-r  & 0  \\
 0 & 0 & 1  
\end{array}
\right).
\label{eq:A_gs_D2BN}
\end{eqnarray}
This is invariant under the D$_{2}$ symmetry.
The generators of the D$_{2}$ group are given by
\begin{eqnarray}
   \bigl\{O\bigr\}
= \bigl\{\vec{1}_{3},I_{1},I_{2},I_{3}\bigr\},
\end{eqnarray}
with $e^{i\alpha}=1$ ($\alpha=0$).
Each element is defined by
\begin{eqnarray}
 \vec{1}_{3}
=
\left(
\begin{array}{ccc}
 1 & 0 & 0  \\
 0 & 1 & 0  \\
 0 & 0 & 1  
\end{array}
\right),
\hspace{0.5em}
 I_{1}
=
\left(
\begin{array}{ccc}
 1 & 0 & 0  \\
 0 & -1 & 0  \\
 0 & 0 & -1  
\end{array}
\right),
\hspace{0.5em}
 I_{2}
=
\left(
\begin{array}{ccc}
 -1 & 0 & 0  \\
 0 & 1 & 0  \\
 0 & 0 & -1  
\end{array}
\right),
\hspace{0.5em}
 I_{3}
=
\left(
\begin{array}{ccc}
 -1 & 0 & 0  \\
 0 & -1 & 0  \\
 0 & 0 & 1  
\end{array}
\right).
\end{eqnarray}
We notice that $\vec{1}_{3}$ is a unit matrix, and $I_{i}$ ($i=1,2,3$) indicate the $\pi$-rotation around the $i$th axis.
It is easily checked that $A_{\mathrm{D}_{2}\mathrm{BN}}(\tau,\vec{x})$ is invariant under the transformation $A_{\mathrm{D}_{2}\mathrm{BN}}(\tau,\vec{x}) \rightarrow O A_{\mathrm{D}_{2}\mathrm{BN}}(\tau,\tilde{\vec{x}}) O^{t}$ ($O \in \mathrm{D}_{2}$).
This phase is called the D$_{2}$-biaxial nematic (D$_{2}$-BN) phase, because the remaining symmetry is the $\mathrm{D}_{2}$ symmetry.

\subsection{D$_{4}$-BN phase ($r=-1$)}

For $r=-1$, the order parameter $A(\tau,\vec{x})$ is expressed as
\begin{eqnarray}
  A_{\mathrm{D}_{4}\mathrm{BN}}(\tau,\vec{x})
=
A_{0}
\left(
\begin{array}{ccc}
 -1 & 0  & 0  \\
 0 & 0  & 0  \\
 0 & 0 & 1  
\end{array}
\right),
\label{eq:tau_gs_D4BN}
\end{eqnarray}
This is invariant under the D$_{4}$ symmetry.
The generators of the D$_{4}$ group are given by
\begin{eqnarray}
   \bigl\{e^{i\alpha},O\bigr\}
=\bigl\{(1,\vec{1}_{3}),\hspace{0.1em}(-1,R_{2}),\hspace{0.1em}(1,I_{2}),\hspace{0.1em}(-1,I_{2}R_{2}),\hspace{0.1em}(1,I_{1}),\hspace{0.1em}(1,I_{3}),\hspace{0.1em}(-1,I_{1}R_{2}),\hspace{0.1em}(-1,I_{3}R_{2})\bigr\},
\end{eqnarray}
where $\vec{1}_{3}$ and $I_{i}$ ($i=1,2,3$) have been defined in the D$_{2}$ group, and $R_{2}$ is defined by
\begin{eqnarray}
 R_{2}
=
\left(
\begin{array}{ccc}
 0 & 0 & -1  \\
 0 & 0 & 0  \\
 1 & 0 & 0  
\end{array}
\right),
\end{eqnarray}
which indicates the $\pi/2$-rotation around the 2nd axis ($y$-axis).
We notice the phase $\bigl\{e^{i\alpha}\bigr\}=\bigl\{1,-1\bigr\}\in \mathbb{Z}_{2}$ ($\alpha=0,\pi$) is locked with the rotation in the spin space and in the real space.
It is easily checked that $A_{\mathrm{D}_{4}\mathrm{BN}}(\tau,\vec{x})$ is invariant under the transformation $A_{\mathrm{D}_{4}\mathrm{BN}}(\tau,\vec{x}) \rightarrow e^{i\alpha} O A_{\mathrm{D}_{4}\mathrm{BN}}(\tau,\tilde{\vec{x}}) O^{t}$ ($(e^{i\alpha},O) \in \mathrm{D}_{4}$).
This phase is called the D$_{4}$-biaxial nematic (D$_{4}$-BN) phase, because the remaining symmetry is the $\mathrm{D}_{4}$ symmetry.

\bibliography{reference}

\end{document}